\documentclass[journal=jpccck,manuscript=article]{achemso}

\usepackage{chemformula} 
\usepackage[T1]{fontenc} 
\usepackage[abbreviations,nonumberlist]{glossaries-extra} 
\usepackage{nameref} 

\author{Oleg Rubel}
\affiliation{Department of Materials Science and Engineering, McMaster University, 1280 Main Street West, Hamilton, Ontario L8S 4L8, Canada}
\email{rubelo@mcmaster.ca}
\author{Xavier Rocquefelte}
\affiliation{Univ Rennes, CNRS, ISCR (Institut des Sciences Chimiques de Rennes) UMR 6226, 263 Av. G{\'e}n{\'e}ral Leclerc, 35700 Rennes, France}

\title{Defect Tolerance of Lead-Halide Perovskite (100) Surface Relative to Bulk: Band Bending, Surface States, and Characteristics of Vacancies}

\makeglossaries
\setabbreviationstyle[acronym]{long-short}
\newacronym{CBE}{CBE}{conduction band edge}
\newacronym{DFT}{DFT}{density functional theory} 
\newacronym{HSE06}{HSE06}{Heyd-Scuseria-Ernzerhof 2006}
\newacronym{MA}{MA}{methylammonium}
\newacronym{PBE}{PBE}{Perdew, Burke, and Ernzerhof}
\newacronym{PR}{PR}{participation ratio}
\newacronym{SI}{SI}{supporting information}
\newacronym{SOC}{SOC}{spin-orbit coupling}
\newacronym{VASP}{VASP}{{V}ienna \textit{ab initio} simulation package}
\newacronym{VBE}{VBE}{valence band edge}
\newacronym{XC}{XC}{exchange-correlation}

\begin{document}

\begin{tocentry}

\includegraphics{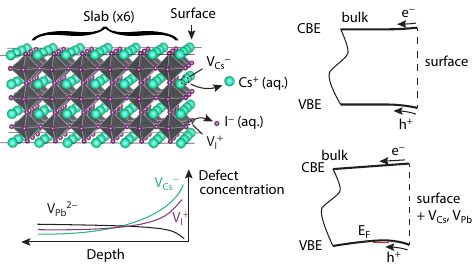}

\end{tocentry}

\begin{abstract}
We characterized the formation of vacancies at a surface slab model and contrasted the results with the bulk of lead-halide perovskites using cubic and tetragonal \ch{CsPbI3} as representative structures. The defect-free \ch{CsI}-terminated (100) surface does not trap charge carriers. In the presence of defects (vacancies), the surface is expected to exhibit $p$-type behavior. The formation energy of cesium vacancies $V_\text{Cs}^{-}$ is lower at the surface than in the bulk, while iodine vacancies $V_\text{I}^{+}$ have a similar energy (around 0.25$-$0.4 eV) within the range of chemical potentials compatible with solution processing synthesis conditions. Lead-iodine divacancies ($V_\text{PbI}^{-}$) are expected to dominate over lead-only vacancies at the surfaces. Major surface vacancies create shallow host-like energy states with a small Franck-Condon shift, making them electronically harmless (same as in bulk). The spin-orbit coupling contributes to the defect tolerance of lead-halide perovskite surfaces by causing delocalization of electronic states associated with $n$-type defects and retraction of lowest unoccupied states from the surface due to a mixing of Pb-$p_{x,y,z}$ orbitals.  These results explain a high optoelectronic performance of two-dimensional structures, nanoparticles, and polycrystalline thin films of lead-halide perovskites despite the abundance of interfaces in these materials.
\end{abstract}



\section{Introduction}

Electronic-structure approaches to modeling defects in solids allow us to predict formation energies of defects, defect concentrations at finite temperatures, and charge-state transition levels taking into account growth conditions and position of the Fermi energy in semiconductors \cite{Lany_MSMSE_17_2009,Freysoldt_RMP_86_2014}. Most of the time, such calculations aim to characterize defects in the bulk of a material, which is modeled with periodic boundary conditions in all directions. Far fewer studies address the calculation of defects at interfaces, e.g., hetero-interfaces \cite{Yang_PCCP_19_2017,Gao_JPCM_33_2020} and surfaces \cite{Komsa_PRL_110_2013, Freysoldt_PRB_102_2020}. The presence of an interface alters characteristics of defects compared to the bulk by either changing the formation energy of defects (e.g., \ch{O^{2-}} vacancy on the MgO(110) surface is 0.2~eV more stable \cite{Freysoldt_PRB_102_2020}), or by introducing new localized states (an N vacancy introduced interface states \cite{Gao_JPCM_33_2020}), or by  changing the defect concentration (segregation of oxygen vacancies \cite{Yang_PCCP_19_2017}).

Single-crystal lead-halide perovskites (organic and inorganic) have a very low concentration of deep electronic traps \cite{Cho_N_14_2022}, which is attributed to the `defect tolerance' of perovskite materials supported by numerous models \cite{Yin_AM_26_2014,Agiorgousis_JACS_136_2014,Yin_APL_104_2014,Du_JPCL_6_2015,Brandt_MC_5_2015,Kang_JPCL_8_2017,Walsh_NM_16_2017,Li_APL_111_2017,Meggiolaro_EES_11_2018}. The `defect tolerance' is associated with shallow transition energy levels of major native point defects in bulk. Interestingly, layered perovskites \cite{Zhang_AM_30_2018}, nanoparticles \cite{Aygueler_JPCC_119_2015,Protesescu_NL_15_2015}, and polycrystalline thin films \cite{Wolf_JPCL_5_2014,Zhao_SA_8_2022} demonstrate good optoelectronic activity. This suggests that surfaces and grain boundaries are also not a significant source of deep electronic traps. Additional efforts are directed towards surface passivation to obtain photovoltaic devices with record-high efficiencies \cite{Lu_iS_23_2020,Park_AEL_7_2022}. Strategies for surface passivation include a reactive post-growth treatment of the surface with larger organic cations to favor $n$-type doping vs typical $p$-type conductivity or intrinsic nature \cite{Jiang_N_611_2022} resulting in a twofold extended lifetime of charge carriers \cite{Chen_NE_8_2023}.

Prior theoretical models \cite{Buin_NL_14_2014,Haruyama_JPCL_5_2014,Meggiolaro_AEL_4_2019} of a \gls{MA} lead iodide (\ch{MAPbI3}) surface arrived at a general conclusion about the absence of deep level surface states within the band gap. However, the literature remains controversial about the localization or delocalization of electronic states at the band edges in slab models. For instance, \citet{Buin_NL_14_2014} reported the delocalization of electronic states at both band edges of a \ch{MAPbI3} slab model simulated with the \gls{DFT}. \citet{Haruyama_JPCL_5_2014} modeled the band edges of a \ch{MAPbI3}(100) stab with terminations and concluded that \gls{VBE} states are localized at the surface, while \gls{CBE} states are delocalized under \ch{PbI} termination conditions. \citet{Meggiolaro_AEL_4_2019} showed the attraction of \gls{CBE} states to the (100) surface of \ch{MAI}-terminated \ch{MAPbI3} and the repulsion of \gls{VBE} states. \citet{Stoumpos_C_2_2017} found that both \gls{VBE} and \gls{CBE} states of a Ruddlesden-Popper bulk phase \ch{(CH3(CH2)3NH3)2(CH3NH3)4Pb5I16} ($n=5$) were strongly localized within one atomic layer near the surface (\gls{VBE} and \gls{CBE} states did not overlap in real space), while states at the band edges were absent within the slab.

The literature on the theoretical characterization of defects at the surface of perovskites is rather scarce and patchy. \citet{Mosconi_EES_9_2016} alluded to a higher concentration of defect complexes (\ch{I_{i}^{-}/$V$_{I}^{+}} Frenkel pairs) at the surface of \ch{MAPbI3} due to a lower formation energy of such defects at \ch{MAI}-vacant or \ch{PbI2}-terminated surfaces but not at the regular \ch{MAI}-terminated surface; the analysis is limited to charge-neutral configurations. \citet{Perez_JPCL_13_2022} recently concluded that 2D Ruddlesden-Popper perovskites (single-layer butylammonium lead iodide) generally retain their defect tolerance with a limited perturbation of the electronic structure (shallow states for iodine vacancy). This conclusion was derived from an analysis of changes in the density of states of structures with iodine, butylammonium, and \ch{PbI2} vacancies as well as iodine interstitials. Only the latter leads to deep localized states within the band gap \cite{Perez_JPCL_13_2022}. To assure the robustness of their conclusions with respect to thermal fluctuations of ionic positions, authors of Ref.~\citenum{Perez_JPCL_13_2022} performed a molecular dynamics simulation of structures with defects up to 2~ps and sampled the electronic structure at several intermediate snapshots. \citet{Brinck_AEL_4_2019} studied neutral defect complexes in \ch{CsPbI3} nanocrystals (non-periodic models) without \gls{SOC} and obtained rather large formation energies (3$-$8~eV) of vacancies without much difference between bulk and the surface location. Most defect complexes that show a negative (or low positive) formation energy also form localized states in the gap \cite{Brinck_AEL_4_2019}, which could be attributed to peculiarities of charge balancing in a defect-free nanoparticle being intrinsically not insulating.

\citet{Song_CMS_194_2021} reported the most comprehensive comparison of defects' characteristics in slab vs bulk, and their approach appears closest to our work. Authors of Ref.~\citenum{Song_CMS_194_2021} studied 12 native point defects in Ruddlesden-Popper butylammonium, \gls{MA} lead iodide ($n=1,2,3$, organic cations were modeled as Cs) at the \gls{DFT} level without \gls{SOC}. Shallow transition energy levels in 2D perovskites appeared deeper in the gap compared to the same levels in the bulk structure, and extended even deeper into the gap as defects approached the surface in 2D structures. Vacancies were the dominant defects with the lowest formation energy in both bulk and in the 2D structure ($n=3$).  Counterintuitively, formation energies of neutral vacancies ($V_\text{Cs}$, $V_\text{Pb}$, and $V_\text{I}$) were 0.2$-$0.7~eV \textit{lower} in bulk than at the surface of the Ruddlesden-Popper 2D layer ($n=3$), which does not support the idea of defects segregating at the surface. It is worth noting that formation energies of the low-energy defects were always negative (regardless of the Fermi energy position or the selection of chemical potentials) for all investigated structures \cite{Song_CMS_194_2021}, which suggests spontaneous formation of defects in large quantities during synthesis.

Here, we perform a study of the most basic intrinsic surface defects (native vacancies) in cubic and tetragonal \ch{CsPbI3} that serve as a computationally more efficient proxy for the structure of hybrid lead-halide perovskites at ambient conditions. We expect that these simple structures share essential properties of defects with more complex hybrid lead-halide perovskites, despite the cubic phase of \ch{CsPbI3} being unstable at room temperature \cite{Trots_JPCS_69_2008}. We calculate the formation energy at various charge states and transition energy levels of these defects by contrasting results for the surface vs the bulk. Since our slab models are equivalent to the 2D Ruddlesden-Popper structure with the number of layers $n=5$ and 6 (tetragonal and cubic, respectively), we anticipate the results to also be transferable to layered perovskites. There are several features that differentiate our work from the prior art: (i) Our choice of chemical potentials of atomic species is tailored to the activity of aqueous ions, in order to best represent solution-processing growth conditions. (ii) We provide quantitative characteristics of the localization of defects' electronic states. (iii) We also elucidate the effect of \gls{SOC} on the formation energy of defects and the location of transition energy levels. We identified defect characteristics and types for which \gls{SOC} can (and cannot) be neglected. (iv) When studying transition energy levels, we differentiate between thermodynamic energy levels and electronic transition levels by imposing limits on the structural relaxation (the Franck-Condon principle \cite{Franck_TFS_21_1926,Condon_PR_28_1926}).

\section{Methods}

Electronic structure calculations were performed using \gls{DFT} \cite{Hohenberg_PR_136_1964,Kohn_PR_140_1965} at the \gls{PBE} \cite{Perdew_PRL_77_1996} level of theory implemented in \gls{VASP} \cite{Kresse_PRB_47_1993,Kresse_CMS_6_1996,Kresse_PRB_54_1996} (version 6.4.2). The \gls{DFT}\,+\,D3 method with Becke-Johnson damping \cite{Grimme_JCP_132_2010,Grimme_JCC_32_2011} was used to capture long-range van der Waals interactions. Projector augmented wave pseudopotentials \cite{Kresse_PRB_59_1999} with 9, 14, and 7 valence electrons were employed for Cs, Pb, and I, respectively. The plane wave cut-off energy was set at the upper limit specified by the pseudopotentials (238~eV). This choice of the cut-off energy gives defect formation energies converged to a precision of 0.1~eV, which was verified using a higher cut-off energy (297~eV) for neutral vacancies in bulk without \gls{SOC}.

The Brillouin zone was sampled using a $\Gamma$-centered grid of $k$ points \cite{Monkhorst_PRB_13_1976} with a mesh density of $2 \times 2 \times 2$ for the bulk $4 \times 4 \times 4$ and $3\sqrt{2} \times 3\sqrt{2} \times 4$ supercells, the density of $2 \times 2 \times 1$ for the slab $3 \times 3 \times 6$ and $2\sqrt{2} \times 2\sqrt{2} \times 5$ supercells, the density of $3 \times 3 \times 1$ and $8 \times 8 \times 1$   for the slab $2 \times 2 \times 3$ and $1 \times 1 \times 6$ supercells, respectively. The $k$ mesh was selected having the zone folding in mind, i.e., the $R(1/2,1/2,1/2)$ point with \gls{CBE} and \gls{VBE} of the primitive cubic Brillouin zone is folded into $\Gamma$ in supercells with even multiplicity and into $M(1/2,1/2,0)$ in the slab model with odd multiplicity in the lateral directions. $\Gamma$ and $M$ points are included in our sampling. We tested convergence of the total energy, defect formation energies, and transition energy levels with respect to the $k$ mesh density and concluded that the $\Gamma$-only calculations (often practiced in literature for large sizes of supercells) lead to inaccurate results. We also tested accuracy of the total energy computed with a tetrahedron integration method vs a Gaussian smearing of 0.01~eV (employed in this work) and found differences immaterial for defect characteristics.

The only dynamically stable structure of \ch{CsPbI3} is the orthorhombic ($\delta$) non-perovskite $P$nma phase \cite{Yang_JCP_152_2020}. However, its electronic structure is very distinct from tetragonal or cubic perovskite structures \cite{Ming_JMCA_10_2022, Sutton_AEL_3_2018}. For this reason, we selected cubic and tetragonal structures of \ch{CsPbI3}, as computationally more efficient representatives of hybrid lead-halide perovskites, such as \ch{(CH3NH3)PbI3}, that has a tetragonal phase very close to cubic at room temperature \cite{Stoumpos_IC_52_2013, Whitfield_SR_6_2016, Zheng_PRM_2_2018}. The initial structure of cubic \ch{CsPbI3} ($P$m$\bar{3}$m, no.~221, $a=6.414$~{\AA}) was obtained from the Materials Project \cite{Jain_AM_1_2013} (ID 1069538). The calculated stress for the structure did not exceed 1~kbar, thus the structure did not require an additional relaxation of the lattice parameters. The initial tetragonal structure of \ch{CsPbI3} ($I$4/mcm, no.~140, $a=8.749$~{\AA} and $c=12.748$~{\AA}) was constructed based on the tetragonal phase of \ch{(CH3NH3)PbI3} \cite{Stoumpos_IC_52_2013} and subsequently relaxed. The stress was also less than 1~kbar in the defect-free supercell and in the slab (in lateral directions), which allows us to fix the lattice and relax only atomic positions (when perturbed by a defect or surface reconstruction) to a desired force threshold. The energy and force convergence thresholds were set at $10^{-7}$~eV and $10^{-2}$~eV~{\AA}$^{-1}$, respectively. A relatively tight force threshold was essential for convergence of the total energies (and defect formation energies, eventually) given the large number of atoms in the supercells.

Calculations were performed with and without \gls{SOC}. When \gls{SOC} was included, the calculation was initialized in a zero spin configuration. A non-spin-polarized calculation mode was selected in the absence of \gls{SOC}. Relaxation of atomic positions was done without \gls{SOC}, and the relaxed structures were used in the calculations with \gls{SOC} without additional force optimization to save computational time. The latter simplification leads to a maximum force still below 0.02$-$0.05~eV~{\AA}$^{-1}$ after adding \gls{SOC}. \gls{HSE06} \cite{Heyd_JCP_118_2003,Krukau_JCP_125_2006} \gls{XC} functional in combination with \gls{SOC} was used for benchmarking purposes on a small slab model to assure the validity of conclusions drawn from \gls{PBE} calculations that suffer from a large band gap error.

Thermodynamic properties of defects were analyzed using the PyDEF~2 code \cite{Pean_CPL_671_2017,Stoliaroff_JCC_39_2018} with minor modifications (see the fork at GitHub \cite{git_PyDEF2_fork}) that enable processing of \gls{VASP} \texttt{OUTCAR} files when \gls{SOC} is included and compatibility with \gls{VASP} version~6. The formation energy of a defect $D$ in a charge state $q$ includes the following terms \cite{Zhang_PRL_67_1991,Lany_PRB_78_2008,Freysoldt_RMP_86_2014}
\begin{equation}\label{eq:Ed}
    E_{\text{d}}(D^q,\Delta E_{\text{F}}) = 
    E_{\text{tot}}(D^q) - E_{\text{tot}}(\text{host})
    \pm \sum_{\alpha} \mu(\alpha)
    + q[E_{\text{VBE}}(\text{host}) + \Delta E_{\text{F}}]
    + E_{\text{corr}},
\end{equation}
where $E_{\text{tot}}$ is the \gls{DFT} total energy, $\mu$ is the chemical potential of each atomic species $\alpha$ added to (negative sign) or subtracted from (positive sign) the host cell, $E_{\text{VBE}}(\text{host})$ is the \gls{VBE} eigenvalue of the host cell, $\Delta E_{\text{F}}$ defines the position of the Fermi energy relative to $E_{\text{VBE}}(\text{host})$, and $E_{\text{corr}}$ is a correction term applied to $E_{\text{tot}}(D^q)$. We included the following corrections: (i) a filling of dispersive states in a finite-size supercell \cite{Lany_PRB_78_2008} (the Moss-Burstein effect), (ii) a potential alignment between the neutral host and the charged cell \cite{Persson_PRB_72_2005}. We found that the inclusion of \citet{Makov_PRB_51_1995} monopole correction for charged bulk cells (\gls{VASP} tag \texttt{LMONO=T}) computed with the dielectric permittivity of 6.3 \cite{Srikanth_PCCP_22_2020} (\gls{VASP} tag \texttt{EPSILON=6.3}) has a minor effect on the total energy of $\pm 1e$ charged bulk structures (0.1~eV or less in the bulk model) but results in over-correction of transition levels for shallow, dispersed host-like defect states, and it was eventually abandoned (see also the discussion in Ref.~\citenum{Lany_PRB_78_2008}).

An electrostatic correction for charged defects in a slab is extensively discussed in the literature \cite{Komsa_PRL_110_2013, Noh_PRB_89_2014, Freysoldt_PRB_97_2018, Silva_PRL_126_2021, Freysoldt_PRB_105_2022}. The corrections aim to compensate for a difference in the electrostatic potential energy $U$ of an isolated charged defect in a slab relative to the electrostatic potential energy of the same defect in a periodic slab model \cite{Komsa_PRB_86_2012}
\begin{equation}\label{eq:dU}
    \delta U(q) = U_\text{iso}(q) - U_\text{per}(q).
\end{equation}
(The potential alignment term is already included in the $E_\text{corr}$ term in Eq.~\eqref{eq:Ed}.) The electrostatic energy resulting from the additional charge $q$ in the periodic model can be expressed as \cite{Komsa_PRB_86_2012}
\begin{equation}\label{eq:Uperiodic}
    U_\text{per} = \frac{1}{2} \int_\text{cell} \delta\phi(\mathbf{r}) \delta\rho(\mathbf{r}) ~ \mathrm{d}\mathbf{r},
\end{equation}
where $\delta\rho(\mathbf{r})$ is the excess charge density and $\delta\phi(\mathbf{r})$ is the electrostatic potential due to the excess charge compatible with periodic boundary conditions and the dielectric function profile imposed by the slab model (see \gls{SI} Fig.~S8). It is expected that the excess charge density distribution fulfills $\int_\text{cell} \delta\rho(\mathbf{r})~\mathrm{d}\mathbf{r}=q$. The periodic electrostatic energy for charged slab models $U_\text{per}$ was calculated using the \texttt{sxdefectalign2d} package \cite{Freysoldt_PRB_97_2018, Freysoldt_PRL_102_2009}. In our case, the defect states are \textit{extended} host-like states and the charge $q$ is delocalized in the lateral plane. Since the electrostatic energy of a charge $q$ is inversely proportional to its spatial extent $\beta$ for either two- or three-dimensional distribution \cite{Ciftja_RP_7_2017,Komsa_PRL_110_2013}, it is proposed to set $U_\text{iso}=0$ in the limit of $\beta_{\parallel}\rightarrow\infty$, consistent with states delocalized in the lateral plane of the slab. This limit was also verified using \texttt{sxdefectalign2d} by increasing the lateral broadening to $\beta_{\parallel}=200$~bohr and observing $\lim_{\beta_{\parallel}\rightarrow\infty} U_\text{iso} = 0$.  For large slab models and the charge state of $q=\pm 1$, relevant for most defects studied here, the correction $\delta U$ amounts to $-0.06 \ldots -0.11$~eV. Since $\delta U$ is a part of the correction term $E_\text{corr}$ in Eq.~\eqref{eq:Ed}, the effect of this correction on the formation energy of charged defects is rather minor (reduction by $-0.06 \ldots -0.11$~eV).

An energy level corresponding to a transition between charge states $q_1$ and $q_2$ in the structure with a defect $D$ is expressed as \cite{Persson_PRB_72_2005}
\begin{equation}\label{eq:trans energy level}
    \varepsilon(D^{q_1/q_2}) =
        \frac{
            E_{\text{d}}(D^{q_1},0) - E_{\text{d}}(D^{q_2},0)
        }{
            q_2 - q_1
        }.
\end{equation}
Calculations of the formation energy of ionized defects can be done either adiabatically (i.e., with the structural relaxation) or following the Franck-Condon principle for electronic transitions (i.e., without the structural relaxation). Here we will present and discuss results obtained with both approaches. The \gls{VASP} output and structure files used in defect calculations are available from the Zenodo repository \cite{Zenodo_10.5281/zenodo.8329348} along with PyDEF~2 output files.

\gls{VASP}-generated \texttt{PROCAR} files were used to analyze the localization of electronic states. The \texttt{PROCAR} output format was modified from the original \gls{VASP} source code to ensure that the output of $k$-, band-, and atom-resolved probability densities had at least three significant digits. The \texttt{prPROCAR.m} Octave script from VASPtools (available on GitHub \cite{Rubel_git_VASPtool}) was employed to extract the $k$- and band-resolved \gls{PR} data.

\section{Results and discussion}

\subsection{Structural model and surface states}

\begin{figure}
  \includegraphics[scale=0.9]{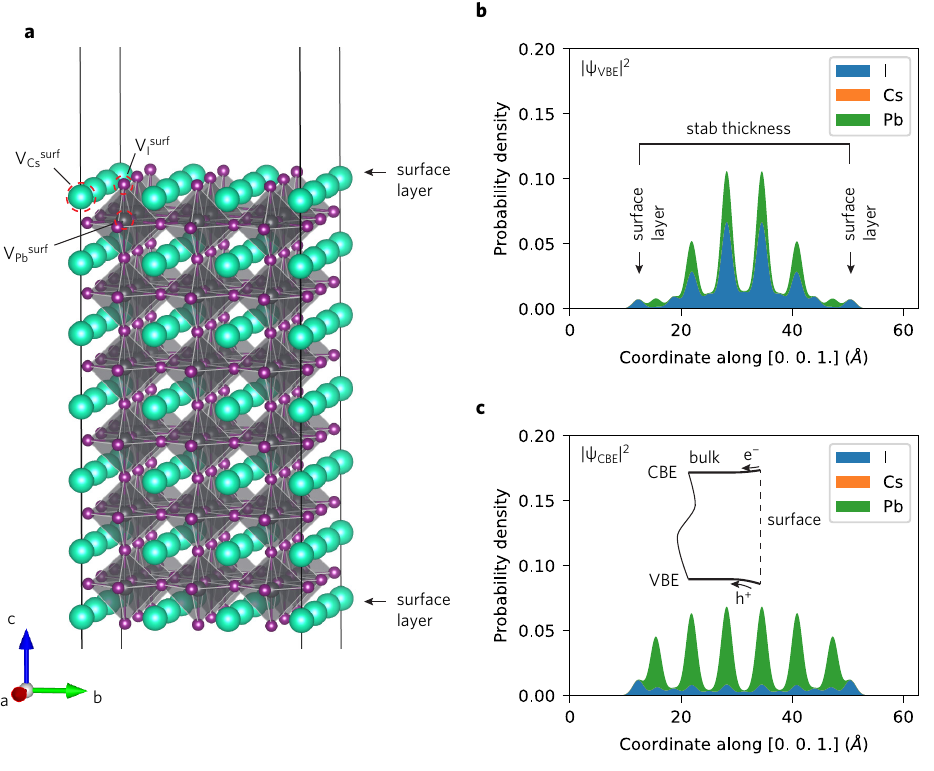}
  \caption{(a) Geometry of a $3 \times 3 \times 6$ cubic \ch{CsPbI3} slab for modeling of (100) surface states. The model is periodic along $\mathbf{a}$ and $\mathbf{b}$ lattice vectors. Location of defects (vacancies) at the surfaces is indicated. (b,c) Spatial distribution of the atom-resolved ($\alpha$) wave function probability density $|\psi_{n,\mathbf{k},\alpha}|^2$ for the band index $n$ and the wave vector $\mathbf{k}$ that correspond to the band edges \gls{VBE} and \gls{CBE} of the slab model. Contribution of individual atoms in a plane with a coordinate $z$ are stacked on top of each other. \gls{SOC} is included. A Gaussian smearing of $\sigma=1$~{\AA} is applied to all atoms.}
  \label{fig-structure}
\end{figure}

An atomistic model of the (100) surface (also employed later for studying defects) is shown in Fig.~\ref{fig-structure}a. The model features a slab with a thickness equivalent to six \ch{[PbI6]} octahedra, \ch{CsI} surface termination, and periodic boundary conditions in two lateral directions. The structure is equivalent to a two-dimensional version of a Ruddlesden-Popper phase \cite{Ruddlesden_AC_11_1958} with the composition \ch{9 Cs7Pb6I19} that falls into a general formula $A_{n+1}B_{n}X_{3n+1}$ with $n=6$. The factor of 9 reflects $3 \times 3$ supercell multipliers in the lateral plane. The vacuum region of ca. 25~{\AA} prevents any spurious interaction between periodic images in the direction $z$ perpendicular to the surface. We deliberately selected the structure with the inorganic cation (\ch{Cs+}) instead of its organic counterparts (e.g., \ch{CH3NH3+}). Organic molecules introduce dipoles that will assume some ordering during the static \gls{DFT} calculation and can induce a large electric field when misoriented or form domains \cite{Frost_NL_14_2014,Liu_JPCL_6_2015}. Under ambient conditions, the fast thermal motion of molecules averages out dipole orientations \cite{Whitfield_SR_6_2016}, which is effectively represented in the static model by a large spherical cation such as cesium. After relaxing the surface structure, we observed a surface reconstruction that mainly consists of attraction of Cs atoms at the surface layer towards the slab center by ca.~0.8~{\AA}.

The lowest band gap of 0.5~eV for the cubic slab model with \gls{SOC} occurs at $\mathbf{k}=(1/2,1/2,0)$, which is expected due to the zone folding and the odd multiplier (3) used in construction of the supercell in the lateral direction. The bulk band gap is only 0.2~eV with \gls{SOC}, which is extremely underestimated relative to the experimental low-temperature band gap of $E_g = 1.72$~eV for cubic \ch{CsPbI3} \cite{Yang_AEL_2_2017}. This shortcoming is usual for \gls{DFT}-\gls{PBE} with \gls{SOC} and this material system. The prior study \cite{Stoumpos_C_2_2017} of $n=5$ Ruddlesden-Popper phase \ch{(CH3(CH2)3NH3)2(CH3NH3)4Pb5I16} also reported a low band gap $E_g \approx 0.3$~eV (\gls{DFT}-\gls{PBE} with \gls{SOC}). Prior calculation \cite{Sutton_AEL_3_2018} of cubic \ch{CsPbI3} bulk quoted $E_g \approx 0.5$~eV (\gls{DFT}-\gls{PBE} with \gls{SOC}) vs $E_g \approx 1.6$~eV (quasiparticle $GW$ with \gls{SOC}). The large size of our supercells for studying defects precludes calculations at higher levels of theory (e.g., hybrid functionals or quasiparticle $GW$) to correct for the band gap error. The band gap error of \gls{DFT}-\gls{PBE} can be largely recovered when \gls{SOC} is omitted as a result of the error cancellation \cite{Even_JPCL_4_2013} ($E_g=1.3$~eV in bulk and $E_g=1.4$~eV in slab). Later we will find this correction essential when calculating the formation energy of $n$-type defects with energy levels close to the \gls{CBE}. Increase of the band gap for the slab model vs bulk can be attributed to a quantum confinement effect.

Since transport of charge carriers occurs at band edges, it is interesting to explore the spatial distribution of states associated with \gls{VBE} and \gls{CBE} in the model with the surface. Figure~\ref{fig-structure}b,c shows atom-resolved $|\psi(\mathbf{r})|^2$ for these two states derived from the \gls{VASP} \texttt{PROCAR} file generated in a calculation with \gls{SOC} for the cubic structure (data for the tetragonal phase are shown in \gls{SI} Fig.~S1). The \gls{VBE} state is confined closer to the center of the slab, while the \gls{CBE} state is distributed more evenly throughout the slab thickness. Notably, band edge states are not localized at the surface in the calculation with \gls{SOC}, which implies that the surface does not act as a trap for free carriers (see the effective band diagram in the inset in Fig.~\ref{fig-structure}c). This result is in line with the earlier findings of \citet{Buin_NL_14_2014}; it is also backed up by calculations at the \gls{HSE06} level of theory (see the \gls{SI} section Fig.~S2a,b) to ensure that this is not an artifact of the \gls{PBE} underestimated band gap. The tetragonal structure favors an even stronger repulsion of holes from the surface, but also a weak attraction of electrons to the surface (see \gls{SI} Fig.~S1).

Our results for $|\psi(\mathbf{r})|^2$ in Fig.~\ref{fig-structure}b,c are similar to the results of \citet{Haruyama_JPCL_5_2014} for \ch{MAPbI3}(100) PbI-terminated surface with \gls{SOC}. However, they are qualitatively different from those by  \citet{Stoumpos_C_2_2017}, who found \gls{VBE} and \gls{CBE} states of a Ruddlesden-Popper bulk phase \ch{(CH3(CH2)3NH3)2(CH3NH3)4Pb5I16} ($n=5$) being confined near the surface. The difference can be attributed (presumably) to peculiarities in the static arrangement of \gls{MA} dipoles having mirror symmetry with respect to the slab center. This could have created a zig-zag potential causing spatial separation between \gls{VBE} and \gls{CBE} states as in Ref.~\citenum{Liu_JPCL_6_2015}. Our structure does not have those polar molecules and should give results relevant to ambient conditions (see the above argument about thermal motion). Experimental data of  \citet{Dymshits_JMCA_2_2014} and \citet{Kim_JPCC_123_2019} give two contradictory pictures of the band bending at the (100) surface of \ch{MAPbI3}.

In the calculation \textit{without} \gls{SOC}, the spatial distribution of the \gls{CBE} state slightly changed its behavior towards attraction to the surface, while the \gls{VBE} state (not affected by \gls{SOC}) remains unchanged (see \gls{SI} Fig.~S3a,b). Data presented by \citet{Meggiolaro_AEL_4_2019} show the same behavior without \gls{SOC} for the (100) surface of \ch{MAI}-terminated \ch{MAPbI3}. \citet{Haruyama_JPCL_5_2014} also noted the same qualitative difference between lowest unoccupied orbitals calculated with and without \gls{SOC} for the (100) \ch{PbI}-terminated surface of \ch{MAPbI3}. This observation suggests that the non-relativistic electronic structure shows less tolerance to surface states. (\citet{Brandt_MC_5_2015} also noted the essential role of \gls{SOC} in the defect tolerance of bulk lead-halide perovskites.) To explain this observation, we recall that the \gls{CBE} in lead-halide perovskites corresponds to a Pb-$p_{1/2}$ split-off band in the presence of \gls{SOC} \cite{Even_JPCL_4_2013}, where Pb-$p_{x,y,z}$ orbitals should be mixed in nearly equal proportions by the \gls{SOC} Hamiltonian. The orbital composition of the \gls{CBE} state of our slab model confirms that expectation
$$
    |\psi_\text{CBE}|^2 = 
    0.31 |p_{x,\text{Pb}}|^2 + 
    0.31 |p_{y,\text{Pb}}|^2 + 
    \mathbf{0.16} |p_{z,\text{Pb}}|^2 + 
    ... \quad \text{(with SOC)}.
$$
The $p_z$ orbital is affected by the quantum confinement of the slab, and we would anticipate it to dominate in the center of the slab following analogy with the `particle in a box' problem. However, the orbital mixing is not required without \gls{SOC}, thus the \gls{CBE} state is comprised of only Pb-$p_{x,y}$ orbitals
$$
    |\psi_\text{CBE}|^2 = 
    0.40 |p_{x,\text{Pb}}|^2 + 
    0.40 |p_{y,\text{Pb}}|^2 + 
    \mathbf{0} |p_{z,\text{Pb}}|^2 + 
    ... \quad \text{(without SOC)}
$$
that are not subjected to confinement in the lateral plane of the slab; the Pb-$p_z$ states are decoupled from Pb-$p_{x,y}$ states and moved to higher energies within the conduction band and assume the shape expected from the quantum confinement.

To elucidate the role of surface reconstruction, we performed a similar analysis of \gls{VBE} and \gls{CBE} states on a model without the relaxation of atomic positions (see \gls{SI} Fig.~S3c,d). The $|\psi(\mathbf{r})|^2$ envelope function profiles for both band edges are nearly identical and closely resemble a `particle in a box' solution. Thus, the surface reconstruction creates an additional electric field pointing from the surface into the slab that shapes \gls{VBE}/\gls{CBE} states, causing an additional repulsion/attraction of $|\psi(\mathbf{r})|^2$ from/to the surface. It should be noted that details of the surface reconstruction are specific to the cation (Cs in this case). The magnitude and direction of the electric field can be influenced by the size of molecules at the surface and the presence of a dipole moment within the molecule \cite{Jiang_N_611_2022}.

\subsection{Formation energy of neutral defects}\label{Sec:Results-Formation-energy}

Here, we will calculate the formation energy of native vacancies according to Eq.~\eqref{eq:Ed} to determine propensity of their occurrence at the surface vs bulk. The formation energy depends on the chemical potential of elements in a reservoir, which is generally expressed as
\begin{equation}\label{eq:mu}
    \mu=\mu^{\circ} + \Delta \mu,
\end{equation}
where $\mu^{\circ}$ is the chemical potential of an element under standard thermodynamic conditions, and $\Delta \mu$ captures any deviations from the standard conditions. For solid elements, it is sufficient to approximate $\mu^{\circ}$ by using the \gls{DFT} total energy $E_{\text{tot}}$ of their most stable polymorph \cite{Zhang_PRL_67_1991}. The chemical potential adjustment ($\Delta \mu$) reflects the chemical activity of elements under a specific synthesis environment and obeys restrictions imposed by the thermodynamic stability of solid phases \cite{Persson_PRB_72_2005}, as well as additional constraints due to the presence of reactants as ions in solution
\begin{subequations}\label{eq:Delta-mu}
    \begin{align}
        \Delta \mu(\ch{Cs}) + \Delta \mu(\ch{Pb}) + 3 \Delta\mu(\ch{I}) & \geq \Delta H_{\text{f}}(\ch{CsPbI3}),  \label{eq:Delta-mu-CsPbI3}\\
        \Delta \mu(\ch{Cs}) + \Delta\mu(\ch{I}) & < \Delta H_{\text{f}}(\ch{CsI}),  \label{eq:Delta-mu-CsI}\\
        \Delta \mu(\ch{Pb}) + 2\Delta\mu(\ch{I}) & < \Delta H_{\text{f}}(\ch{PbI2}),  \label{eq:Delta-mu-PbI2}\\
        \Delta \mu(\ch{Cs}) & \leq \Delta G^{\circ}(\ch{Cs^{+},~aq}),  \label{eq:Delta-mu-Cs+}\\
        \Delta \mu(\ch{Pb}) & \leq \Delta G^{\circ}(\ch{Pb^{2+},~aq}),  \label{eq:Delta-mu-Pb2+}\\
        \Delta \mu(\ch{I}) & \leq \Delta G^{\circ}(\ch{I^{-},~aq}).  \label{eq:Delta-mu-I-}
    \end{align}
\end{subequations}
The formation free energy of a solid is approximated by enthalpy ($\Delta H_{\text{f}}$) at 0~K in our calculations. It can be noted that the final temperature correction is computationally very expensive, while its effect on the formation enthalpy of solids is much smaller than the chemical accuracy of \gls{DFT}-\gls{PBE} \cite{Jackson_PRB_88_2013}.

Table~\ref{tbl:enthalpies} summarizes \gls{DFT} formation enthalpies of relevant solids compared to the experiment to show the chemical accuracy of our calculations. The largest error is about 0.2~eV per atom. The effect of \gls{SOC} is notable for lead-containing structures without a clear trend on whether the relativistic effects improve or not the agreement with the experiment. The values of $\Delta\mu$ will be tailored to the activity of ionic species in an aqueous solution. From the Pourbaix atlas \cite{Pourbaix1974}, we find
$\Delta G^{\circ}(\ch{Cs^{+},~aq})=-2.92$~eV, 
$\Delta G^{\circ}(\ch{Pb^{2+},~aq})=-0.25$~eV, and 
$\Delta G^{\circ}(\ch{I^{-},~aq})=-0.54$~eV.

\begin{table}
  \caption{Formation enthalpies of solids $\Delta H_{\text{f}}$ (eV/f.u.).}
  \label{tbl:enthalpies}
  \begin{tabular}{lccc}
    \hline
    Method  &  CsI & \ch{PbI2} & \ch{$\delta$-CsPbI3} \\
    &  (Pm$\bar{3}$m, 221)   & (R$\bar{3}$m, 166) & (Pnma, 62) \\
    \hline
    \gls{PBE}+D3 without \gls{SOC} & $-$3.23& $-$1.96 & $-$5.44 \\
    \gls{PBE}+D3 with \gls{SOC} & $-$3.20& $-$1.79 & $-$5.23 \\
    \gls{HSE06} without \gls{SOC} & $-$3.31& $-$2.16& $-$5.66\\
    \gls{HSE06} with \gls{SOC} & $-$3.29& $-$1.98& $-$5.46\\
    Experiment & $-$3.61 \cite{Cordfunke_TA_90_1985}& $-$1.81 \cite{Chase_JPCRD_Monograph9_1998} & $-$5.60\textsuperscript{\emph{a}}\\
    \hline
  \end{tabular}
  \\
  \raggedright
  \textsuperscript{\emph{a}} Estimated based on the experimental enthalpy of $-0.18$~eV/f.u. for the reaction \ch{CsI + PbI2 -> CsPbI3} (Ref.~\citenum{Wang_JACS_141_2019}).
\end{table}

\begin{figure}
  \includegraphics{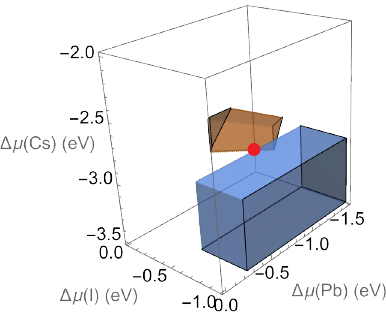}
  \caption{Region of \ch{CsPbI3} thermodynamic stability (brown, based on \gls{DFT} data without \gls{SOC}) that satisfies Eqs.~\eqref{eq:Delta-mu-CsPbI3}-\eqref{eq:Delta-mu-PbI2} and the region of chemical potentials of ions accessible in an aqueous environment (blue) that satisfies Eqs.~\eqref{eq:Delta-mu-Cs+}-\eqref{eq:Delta-mu-I-}. The red dot represents the point of overlap between the two regions.}
  \label{fig-chem-pot}
\end{figure}

Figure~\ref{fig-chem-pot} illustrates the region of chemical potentials corresponding to the stability of the perovskite phase. At the same time, we need to acknowledge that the energy of dissolved species is additionally lowered relative to their bulk phase by $\Delta\mu$: the more soluble the ion is, the lower $\Delta\mu$ is (e.g., \ch{Cs+}). Conditions that favor \ch{CsPbI3} precipitation from the solution (red point in Fig.~\ref{fig-chem-pot}) allow us to fix $\Delta\mu$
\begin{subequations}\label{eq:Delta-mu-solved}
    \begin{align}
        \Delta \mu(\ch{Cs}) & \approx \Delta G^{\circ}(\ch{Cs^{+},~aq}),  \label{eq:Delta-mu-Cs-solved}\\
        \Delta \mu(\ch{I}) & \approx \Delta G^{\circ}(\ch{I^{-},~aq}),  \label{eq:Delta-mu-I-solved}\\
        \Delta H_{\text{f}}(\ch{CsPbI3}) - \Delta G^{\circ}(\ch{Cs^{+},~aq}) - \ldots & \nonumber\\
        \ldots 3\Delta G^{\circ}(\ch{I^{-},~aq}) \leq \Delta \mu(\ch{Pb}) & \leq \Delta H_{\text{f}}(\ch{PbI2}) -2\Delta G^{\circ}(\ch{I^{-},~aq}).  \label{eq:Delta-mu-Pb2-solved}
    \end{align}
\end{subequations}
Results for the chemical potentials are summarized in Table~\ref{tbl:chem potentials}. The standard conditions at which $\Delta G^{\circ}$'s are derived assume a 1~M concentration of aqueous species, while the chemical potentials can be additionally adjusted in case of significant deviations from that (see p.~5-113 in Ref.~\citenum{Haynes_CRC-handbook-chem-phys-ed97th}). We have two sets of chemical potentials: with and without \gls{SOC}. The effect of \gls{SOC} on the chemical potential is largest for lead, as it is the heaviest element.

\begin{table}
  \caption{Chemical potentials of elements that fulfill constraints \eqref{eq:Delta-mu-solved} relevant to solution-processed synthesis. Two sets of data represent different levels of theory: \gls{PBE}\,+\,D3 and \gls{HSE06}.}
  \label{tbl:chem potentials}
  \begin{tabular}{lcccc}
    \hline
    Element  &  \multicolumn{2}{c}{Without \gls{SOC}} & \multicolumn{2}{c}{With \gls{SOC}}\\
    & $\Delta\mu$~(eV) & $\mu$~(eV) & $\Delta\mu$~(eV) & $\mu$~(eV)\\
    \hline
    \gls{PBE}\,+\,D3\\
    \quad Cs & $-$2.92 & $-$3.909 & $-$2.92 & $-$4.050 \\
    \quad Pb & $-$0.89 & $-$4.724 & $-$0.70\textsuperscript{\emph{a}} & $-$5.044 \\
    \quad I  & $-$0.54 & $-$2.252 & $-$0.54 & $-$2.409 \\
    \hline
    \gls{HSE06}\\
    \quad Cs & $-$2.92 & $-$3.911 & $-$2.92 & $-$4.048 \\
    \quad Pb & $-$1.10 & $-$5.238 & $-$0.91 & $-$5.535 \\
    \quad I  & $-$0.54 & $-$2.587 & $-$0.54 & $-$2.732 \\
    \hline
  \end{tabular}
  \\
  \raggedright
  \textsuperscript{\emph{a}} The constraint \eqref{eq:Delta-mu-Pb2-solved} is satisfied with a small error of 0.02~eV.
\end{table}

Our approach to finding the chemical potentials augments the common practice \cite{Yin_APL_104_2014}, which involves defining a region of chemical potentials in which \ch{MAPbI3} is stable, while \ch{MAI} and \ch{PbI2} are not (the brown region in Fig.~\ref{fig-chem-pot}), and selecting two extreme sets of chemical potentials (referred to as I-rich/Pb-poor and I-poor/Pb-rich sets) from the multitude of possibilities \cite{Agiorgousis_JACS_136_2014}. The resultant defect formation energies typically varied by 1$-$3~eV, depending on which extreme set of chemical potentials is selected \cite{Yin_APL_104_2014,Li_APL_111_2017,Song_CMS_194_2021}, which is the largest source of uncertainty in the calculation of $E_{\text{d}}$.

Now we present in Table~\ref{tbl:Ed-neutral} results for the formation energy of neutral defects at the surface vs bulk calculated with and without \gls{SOC} in the cubic structure (large model size). There are several notable points: (i) The surface favors the formation of Cs vacancies. (ii) Lead vacancies are more favorable in the bulk than at the surface. (iii) The formation energies of iodine and lead vacancies are reduced after adding \gls{SOC}. The \gls{SOC} renormalization of the $V_{\text{I}}$ defect formation energy can be attributed to the band gap error (see \gls{SI} Fig.~S4). The iodine vacancy creates an $n$-type defect with one electron promoted to the conduction band. The energy penalty is less when the band gap is underestimated. Thus, the $E_{\text{d}}$ results at \gls{PBE} level without \gls{SOC} are more reliable for $V_{\text{I}}$ (see Table~\ref{tbl:Ed-neutral}) and for other $n$-type defects, in general. However, we should not expect the band gap error to explain the \gls{SOC} renormalization of the $V_{\text{Pb}}$ defect formation energy. The lead vacancy removes two electrons from the valence band, the energy position of which is not much affected by \gls{SOC}. In this case, the renormalization is due to the lower chemical potential of lead with \gls{SOC} (see Table~\ref{tbl:chem potentials}). Thus, the results with \gls{SOC} are more trustworthy for $V_{\text{Pb}}$. The higher energy of the Pb vacancy at the surface can be explained by the very unfavorable position of the iodine atom immediately above the vacancy at the surface, which is nearly detached (see \gls{SI} Fig.~S5b). We tried removing this iodine atom to create a $p$-type $V_\text{PbI}$ divacancy at the surface (see \gls{SI} Fig.~S5d). Its formation energy in the neutral state comes second lowest at the surface after $V_\text{Cs}$ suggesting that the lead-iodine divacancy is more plausible at the surface than the lead vacancy alone. A lower energy of $V_\text{Cs}$ at the surface can be connected with a higher Madelung energy of the surface site facing vacuum and the purely ionic nature of the \ch{Cs-PbI3} bond.

\begin{table}
  \caption{Formation energy (eV) of neutral vacancies in \ch{CsPbI3} calculated at the surface of a slab model and in bulk at two levels of theory (\gls{HSE06} and \gls{PBE}\,+\,D3) and varying sizes of the model for cubic and tetragonal phases. The numerator/denominator correspond to the results obtained without/with \gls{SOC}.}
  \label{tbl:Ed-neutral}
  \begin{tabular}{llllcccc}
    \hline
    Location & Model size & Phase & \gls{XC} approx. &  $V_{\text{Cs}}$   & $V_{\text{I}}$ & $V_{\text{Pb}}$ & $V_{\text{PbI}}$ \\
    \hline
    Slab, surf.   & $3 \times 3 \times 6$ & cubic & \gls{PBE}+D3  & 0.40/0.22 & 1.7/0.88 & 1.4/1.2 & 0.85/0.75 \\
    & $2\sqrt{2} \times 2\sqrt{2} \times 5$ & tetragonal & \gls{PBE}+D3  & 0.48/0.29 & 2.0/1.2 & 1.5/1.2 & 0.88/0.74 \\
    & $2 \times 2 \times 3$ & cubic & \gls{PBE}+D3  & 0.54/0.36 & 1.8/1.1 & 1.7/1.5 & --- \\
    & $2 \times 2 \times 3$ & cubic & \gls{HSE06}  & 0.61/0.44 & 1.9/1.2 & 2.2/1.9 & --- \\
    \hline
    Bulk   & $4 \times 4 \times 4$ & cubic & \gls{PBE}+D3  & 0.91/0.74 & 1.6/0.62 & 0.94/0.69 & --- \\
    & $3\sqrt{2} \times 3\sqrt{2} \times 4$ & tetragonal & \gls{PBE}+D3  & 0.95$^a$/0.77 & 1.7$^b$/0.90 & 1.3$^c$/1.0 & --- \\
    \hline
  \end{tabular}
  \\
  \raggedright
  Other calculations for \ch{MAPbI3} and \ch{CsPbI3} including adjustment for differences in the chemical potentials of elements: $^a$~0.77 \cite{Yin_APL_104_2014}, 0.05 \cite{Song_CMS_194_2021} eV, $^b$~1.3 \cite{Yin_APL_104_2014}, 0.96 \cite{Song_CMS_194_2021} eV, $^c$~1.8 \cite{Yin_APL_104_2014}, 0.90 \cite{Song_CMS_194_2021} eV.
\end{table}

Next, we discuss similarities and differences in the formation energy of neutral vacancies in the tetragonal vs cubic structure of \ch{CsPbI3} (Table~\ref{tbl:Ed-neutral}). The tetragonality of \ch{CsPbI3} is close to that of \ch{MAPbI3} at about 150-200~K \cite{Zheng_PRM_2_2018}. At the surface, the defect formation energies of cesium and lead vacancies are not sensitive (within 0.1~eV) to the choice of the phase. The iodine vacancy shows about 0.3~eV greater formation energy in the tetragonal phase, which is largely due to the greater band gap of the tetragonal phase (see \gls{SI} Fig.~S4). The net difference in $E_\text{d}(V_\text{I}^0)$ between the tetragonal structure and the cubic phase (excluding effect of the band gap) is only 0.1~eV. Thus, the formation energies of surface vacancies calculated using the cubic phase also apply to the tetragonal structure. In bulk, the above discussion applies to cesium and iodine vacancies. The formation energy of a lead vacancy indeed shows a greater disparity of ca.~0.3~eV between the cubic and tetragonal structures. To rule out a possible nonphysical structure relaxation \cite{Wu_CPL_39_2022} due to dynamical instability of the cubic structure, we placed the Pb atom back into the relaxed structure with the vacancy and were able to recover (after subsequent structural relaxation) the total energy of the defect-free structure within 1~meV.

In order to assess the sensitivity of defect formation energies to the selection of exchange-correlation functional, we performed additional calculations using a hybrid functional \gls{HSE06} with and without \gls{SOC}. Due to the computational intensity of \gls{HSE06}, we had to use a smaller slab model ($2 \times 2 \times 3$).  Results for the formation energy of neutral vacancies at the surface are listed in Table~\ref{tbl:Ed-neutral}. We also obtained \gls{PBE}\,+\,D3 results for the same model size to facilitate the comparison. The formation energies of a cesium surface vacancy are identical between \gls{PBE}\,+\,D3 and \gls{HSE06} within 0.1~eV. The formation energy of a neutral surface iodine vacancy also seems to be identical between \gls{PBE}\,+\,D3 and \gls{HSE06}, but a more careful analysis taking into account its sensitivity to the calculated band gap (see SI Fig.~S4) shows $E_\text{d}(V_\text{I}^0)$ being 0.2$-$0.3~eV lower at the \gls{HSE06} level of theory for the same band gap. Taking into account the limited model size effect and the correct experimental band gap of about 1.5~eV, we can estimate $E_\text{d}(V_\text{I}^0) \approx 1.5$~eV. The formation energies of neutral lead vacancies differ more significantly (by 0.4$-$0.5~eV) between \gls{HSE06} and \gls{PBE}\,+\,D3, making this defect even more unlikely at the surface.

For completeness, we list the calculated formation energies of vacancies in lead-halide perovskites available in the literature. However, the data scattering is too large to draw any conclusions. The footnotes of Table~\ref{tbl:Ed-neutral} contain data for bulk \ch{MAPbI3} that can be directly compared with our results, since the authors \cite{Yin_APL_104_2014,Song_CMS_194_2021} reported the chemical potentials, and we were able to adjust for the differences with our choice. Other data in the literature show even more variability in the formation energies of neutral vacancies due to a wide range of chemical potentials. \citet{Li_APL_111_2017} reported the energies for \ch{CsPbI3}: $E_\text{d}(V_{\text{Cs}})=1.3\ldots0.3$~eV, $E_\text{d}(V_{\text{I}})=-0.2\ldots0.8$~eV, $E_\text{d}(V_{\text{Pb}})=0.4\ldots-1.5$~eV (the range is from lead-rich to lead-poor conditions). The data from \citet{Buin_NL_14_2014} for defects in bulk \ch{MAPbI3} are: $E_\text{d}(V_{\text{MA}})=1.8\ldots0.7$~eV, $E_\text{d}(V_{\text{I}})=0.7\ldots1.7$~eV, $E_\text{d}(V_{\text{Pb}})=2.7\ldots0.6$~eV.

\subsection{Charged defects and transition energy levels}

We assess the position of energy levels associated with defects relative to the band edges of the defect-free slab model using a transition energy level technique. The goal of this analysis is to find out if electronic states associated with vacancies at the surface of lead-halide perovskites emerge in the band gap (especially deep levels) as those will have an adverse effect on transport properties and can serve as non-radiative recombination centers. As a reference, it is useful to establish the energy levels associated with the creation of an electron or a hole in the defect-free structure, as previously suggested by \citet{Hellstroem_JCTC_9_2013}. Following Eq.~\eqref{eq:trans energy level}, the corresponding energy levels are \cite{Persson_PRB_72_2005}
\begin{subequations}\label{eq:mu_v, mu_c}
    \begin{align}
        \mu_{e,\text{v}} & \equiv \varepsilon(\text{host}^{+/0}) = E_{\text{tot}}(\text{host}) - E_{\text{tot}}^{*}(\text{host}^{+}), \\
        \mu_{e,\text{c}} & \equiv \varepsilon(\text{host}^{0/-}) = E_{\text{tot}}^{*}(\text{host}^{-}) - E_{\text{tot}}(\text{host}).
    \end{align}
\end{subequations}
Here, the asterisk ($^{*}$) implies that the \gls{DFT} total energy of a charged cell includes the correction term $E_{\text{corr}}$ in Eq.~\eqref{eq:Ed}. The total energy difference between the neutral and charged model ($\pm 1e$) is related to the first ionization potential and the electron affinity \cite{Perdew_PRL_49_1982,Schulte_JPCSSP_7_1974,Janak_PRB_18_1978}. For extended states in the dilute limit of hole/electron concentration, $\mu_{e,\text{v/c}}$ approaches the Kohn-Sham \gls{VBE}/\gls{CBE} eigenenergies \cite{Persson_PRB_72_2005}, respectively. Data in Table~\ref{tbl:Transition-ene-levels} show a nearly perfect agreement between $\varepsilon(\text{host}^{\pm/0})$ states and the band edges in the bulk, which indicates the efficiency of corrections to the total energy of charged bulk cells (with the band filling correction being dominant in this case). Without corrections, those levels would be located 0.1$-$0.2~eV away from the band edges, delving into the valence or conduction band. \citet{Gerstmann_PBCM_340–342_2003} also noted that the Makov-Payne monopole term cancels out in the limit of delocalized states.

\begin{table}
  \caption{Thermodynamic transition energy levels (eV) associated with vacancies in cubic and tetragonal \ch{CsPbI3}. Transition levels of the host without defects indicate the positions of perceived band edges. Energies of acceptor/donor-like states are given relative to Kohn-Sham \gls{VBE}/\gls{CBE} eigenvalues.}
  \label{tbl:Transition-ene-levels}
  \begin{tabular}{llccccc}
    \hline
    Structure & State  &  \multicolumn{2}{c}{Surface}   & & \multicolumn{2}{c}{Bulk} \\
    \cline{3-4}\cline{6-7}
    && without \gls{SOC} & with \gls{SOC} & & without \gls{SOC} & with \gls{SOC} \\
    \hline
    Cubic &$\text{host}^{+/0}$   & $E_{\text{VBE}}+0.08$ & $E_{\text{VBE}}+0.07^a$ & & $E_{\text{VBE}}+0.01$ & $E_{\text{VBE}}+0.01$ \\
    &$\text{host}^{0/-}$   & $E_{\text{CBE}}-0.07$ & $E_{\text{CBE}}-0.07^b$ & & $E_{\text{CBE}}-0.01$ & $E_{\text{CBE}}-0.01$ \\
    \cline{2-7}
    &$V_{\text{Cs}}^{0/-}$   & $E_{\text{VBE}}-0.02$ & $E_{\text{VBE}}-0.05^c$ & & $E_{\text{VBE}}+0$ & $E_{\text{VBE}}+0$ \\
    &$V_{\text{I}}^{+/0}$ & $E_{\text{CBE}}-0.02$ & $E_{\text{CBE}}+0.07^d$ & & $E_{\text{CBE}}-0.03$ & $E_{\text{CBE}}+0.02$ \\
    &$V_{\text{Pb}}^{0/2-}$  & --- & $E_{\text{VBE}}+0^e$ & & $E_{\text{VBE}}+0$ & $E_{\text{VBE}}+0$ \\
    &$V_{\text{Pb}}^{0/-}$  & $E_{\text{VBE}}+0.06$ & --- & & --- & --- \\
    &$V_{\text{Pb}}^{-/2-}$  & $E_{\text{VBE}}+0.08$ & --- & & --- & --- \\
    &$V_{\text{PbI}}^{0/-}$  & $E_{\text{VBE}}-0.01$ & $E_{\text{VBE}}-0.04$& & --- & --- \\
    \hline
    Tetragonal & $\text{host}^{+/0}$   & $E_{\text{VBE}}+0.02$& $E_{\text{VBE}}+0.01$& & $E_{\text{VBE}}+0.01$ & $E_{\text{VBE}}+0.01$ \\
    &$\text{host}^{0/-}$   & $E_{\text{CBE}}-0.07$ & $E_{\text{CBE}}-0.05$& & $E_{\text{CBE}}-0.01$ & $E_{\text{CBE}}-0.01$ \\
    \cline{2-7}
    &$V_{\text{Cs}}^{0/-}$   & $E_{\text{VBE}}+0.08$& $E_{\text{VBE}}+0.05$& & $E_{\text{VBE}}+0$ & $E_{\text{VBE}}+0$ \\
    &$V_{\text{I}}^{+/0}$ & $E_{\text{CBE}}-0.19$& $E_{\text{CBE}}-0.04$& & $E_{\text{CBE}}+0$& $E_{\text{CBE}}+0$\\
    &$V_{\text{Pb}}^{0/2-}$  & --- & ---& & $E_{\text{VBE}}-0.01$& $E_{\text{VBE}}+0$ \\
    &$V_{\text{Pb}}^{0/-}$  & $E_{\text{VBE}}+0.07$& $E_{\text{VBE}}+0.07$& & ---& --- \\
    &$V_{\text{Pb}}^{-/2-}$  & $E_{\text{VBE}}+0.48$& $E_{\text{VBE}}+0.36$& & --- & --- \\
    &$V_{\text{PbI}}^{0/-}$  & $E_{\text{VBE}}+0.11$& $E_{\text{VBE}}+0.06$& & --- & --- \\
    \hline
  \end{tabular}
  \\
  \raggedright
  Transition energy levels evaluated at the \gls{HSE06} with \gls{SOC} level of theory for slab model size of $2 \times 2 \times 3$, cubic phase: $^a~E_{\text{VBE}}+0.11$; $^b~E_{\text{CBE}}-0.20$; $^c~E_{\text{VBE}}-0.09$; $^d~E_{\text{CBE}}+0.05$; $^e~E_{\text{VBE}}+0.02$.
\end{table}

The agreement between the band edges and $\varepsilon(\text{host}^{\pm/0})$ transition energy levels is less favorable for the slab model (Table~\ref{tbl:Transition-ene-levels}). Taking the $\varepsilon(\text{host}^{0/-})$ level in the tetragonal slab without \gls{SOC} as an example, we find from uncorrected total energies $\mu_{e,\text{c}}=E_{\text{CBE}}+0.06$~eV. The potential alignment ($-0.08$~eV) and the band filling ($-0.04$~eV) corrections bring the energy level at $\mu_{e,\text{v}}=E_{\text{CBE}}-0.07$~eV indicating some over-correction in the slab model. We performed additional checks and ruled out the $k$-mesh density and the structure relaxation (vertical vs non-vertical transitions) as possible causes for the 0.07~eV discrepancy. Thus, the remaining misalignment between $\mu_{e,\text{v/c}}$ and \gls{VBE}/\gls{CBE} indicates the magnitude of ambiguity that can still be present in the analysis of defect energy levels of the slab model.

An energy correction $\delta U(q)$ associated with the spurious electrostatic energy of charged stacked slabs \cite{Komsa_PRL_110_2013, Noh_PRB_89_2014, Freysoldt_PRB_97_2018, Silva_PRL_126_2021, Freysoldt_PRB_105_2022} was evaluated to determine its ability to explain the slight discrepancy between  the electron chemical potential $\mu_{e,\text{v/c}}$ and \gls{VBE}/\gls{CBE}. In \gls{SI}, we provided details on the calculation of the $U_\text{per}$ for the defect-free tetragonal slab with a charge of $q=\pm 1$ (see \gls{SI} Figs.~S9 and S10). The electrostatic potential energy of defect-free charged stacked slabs is approximately 0.06~eV, regardless of the sign of $q$. Subtracting this energy from the total energy of the charged slabs $E_{\text{tot}}^{*}(\text{host}^{\pm})$ effectively stabilizes the charged state and shifts the transition energy levels $\varepsilon(\text{host}^{+/0})$ and $\varepsilon(\text{host}^{0/-})$ further away from the band edges into the band gap by 0.06~eV enhancing the discrepancy between $\mu_{e,\text{v/c}}$ and \gls{VBE}/\gls{CBE} eigenvalues. It is worth noting that the electrostatic correction in the literature \cite{Komsa_PRL_110_2013, Noh_PRB_89_2014, Freysoldt_PRB_97_2018, Silva_PRL_126_2021, Freysoldt_PRB_105_2022} has always been tested for localized defect states. \citet{Komsa_PRB_86_2012} expressed reservations about the applicability of ``simple electrostatics'' in cases where charge delocalization is observed, as in our case of extended \gls{VBE} and \gls{CBE} states. Therefore, the electrostatic correction is not included in Table~\ref{tbl:Transition-ene-levels}, however its effect can be inferred from Fig.~S11 (see \gls{SI} section) where we presented both corrected and uncorrected transition energy levels for vacancies in slabs of the tetragonal structure.

Table~\ref{tbl:Transition-ene-levels} summarizes transition energy levels for vacancies at the surface, calculated with and without SOC. These values reflect \textit{thermodynamic} energy levels, i.e., they include relaxation of the structures in all charged states \cite{Lyons_APL_97_2010,Alkauskas_JAP_119_2016}. All bulk states are shallow, and their energy position is given relative to the \gls{VBE} or \gls{CBE} energy eigenvalues for $p$- and $n$-type defects, respectively. Most defect energy levels at the surface remain close to the band edges and do not form detrimental deep traps within the band gap, indicating that the tolerance of the electronic structure of lead-halide perovskites to defects also extends to the surface. This conclusion is robust with respect to selection of the exchange-correlation functional (see \gls{HSE06} data in the footnotes of Table~\ref{tbl:Transition-ene-levels}) and supports prior studies by \citet{Song_CMS_194_2021} and \citet{Perez_JPCL_13_2022} based on thinner slab models. The only exception is the deep $\varepsilon(V_\text{Pb}^{-/2-})$ transition energy level at the surface of the tetragonal phase. However, the formation energies (Table~\ref{tbl:Ed-neutral}) suggest that lead-iodine divacancies are significantly more prevalent than lead-only vacancies.

Our data in Table~\ref{tbl:Transition-ene-levels} show that the transition energy levels of vacancies are shallow with \gls{SOC} and remain the same without \gls{SOC}. This is different from the statement by \citet{Brandt_MC_5_2015}, suggesting that the vacancy-type defects in \ch{MAPbI3} are resonant with the band edges due to relativistic effects shifting the \gls{CBE} down in energy. We observe that the $V_{\text{I}}^{+/0}$ energy level shifts up relative to the \gls{CBE} in both slab and bulk models, but the magnitude (0.05$-$0.09~eV) is very small in comparison to the band gap renormalization due to \gls{SOC}. \citet{Lany_PRB_78_2008} noted that this type of defects falls into the category of delocalized host states perturbed by a screened Coulomb potential of a defect, resulting in a shallow energy level that moves together with the associated band edge when the gap changes, e.g., when adding \gls{SOC} or using a high-level theory to circumvent the \gls{DFT} band gap problem. Hence, the energy levels follow the band edges as the band gap opens significantly in calculations without \gls{SOC}.  The more delocalized nature of the $V_{\text{I}}^{+/0}$ state with \gls{SOC} (similar to the discussion on surface states and driven by the Pb-$p_{x,y,z}$ orbital mixing) explains the shallower energy position of the states in both slab and bulk models. (A more quantitative discussion of localization follows in the next subsection.) If we take into account the electrostatic potential energy for charged periodic slabs (see \gls{SI} section for details), the transition energy levels will shift even closer to the band edges (Fig.~S11), reinforcing the shallow nature of charge transition energy level associated with vacancies at the surface.

It should be noted that the energy position of transition levels in lead-halide perovskites is much better reproducible across the literature \cite{Agiorgousis_JACS_136_2014,Yin_APL_104_2014,Buin_NL_14_2014,Yin_AM_26_2014,Meggiolaro_EES_11_2018} than the formation energy of defects. The only notable outliers are the work by \citet{Du_JPCL_6_2015} and by \citet{Meggiolaro_EES_11_2018}. \citet{Du_JPCL_6_2015} found that energy levels of host-like states, such as $\varepsilon(V_{\text{I}}^{+/0})$, do not follow the band edges as if they were localized deep-level states. \citet{Meggiolaro_EES_11_2018} obtained mid-gap states for $V_{\text{Pb}}$ in bulk.

\begin{figure}
  \includegraphics{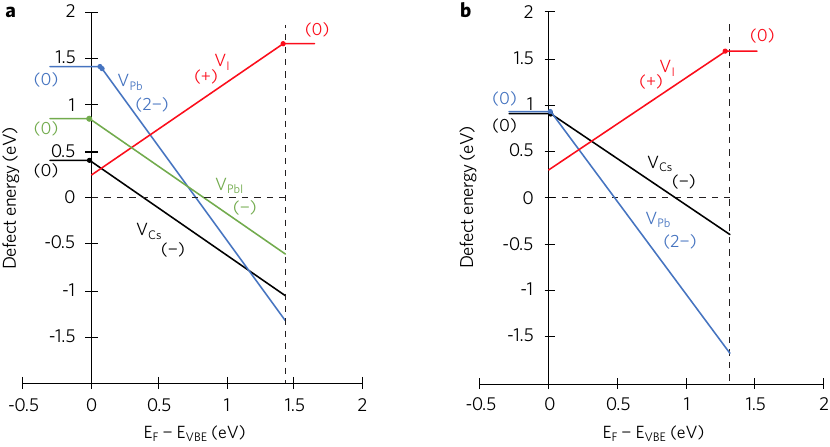}
  \caption{Formation energy of vacancies at the surface (a) and in the bulk (b) of cubic \ch{CsPbI3} as a function of the Fermi energy ($E_{\text{F}}$). Values in brackets represent the charge state. The vertical dashed lines mark the location of the \gls{CBE}. The chemical potential of elements is tuned to reflect solution-processed synthesis conditions. Results are shown without \gls{SOC} to compensate for the \gls{DFT} band gap error. All defect energies associated with the lead vacancy should be adjusted by ca. 0.2~eV (shifted down) to compensate for the overestimation of its chemical potential due to the omission of \gls{SOC}.}
  \label{fig-def-ene}
\end{figure}

Having calculated transition energy levels and formation energy of neutral defects, it is possible to visualize the formation energy of charged defects with respect to the Fermi energy. Results for the cubic phase are presented in Fig.~\ref{fig-def-ene}. Since data for the tetragonal phase are very similar (see \gls{SI} section Fig.~S11), we will focus the discussion on the cubic phase in the remaining part of the paper. In bulk, the Fermi energy is pinned slightly above the \gls{VBE}, which is consistent with the intrinsic or weak $p$-type character of lead-halide perovskites \cite{Dymshits_JMCA_2_2014,Xie_AEM_7_2021,Jiang_N_611_2022,PenaCamargo_APR_9_2022}. Predominant vacancies in bulk are $V_\text{Pb}^{2-}$ and $V_\text{I}^{+}$. (\citet{Yin_APL_104_2014} previously identified $V_\text{Pb}$ as a dominant $p$-type defect in bulk based on \gls{DFT} calculations.)  At the surface, our results suggest pinning of the Fermi energy at the \gls{VBE} and segregation of $V_\text{Cs}^{-}$ and $V_\text{I}^{+}$ vacancies. The ratio $[V_\text{Cs}^{-}]/[V_\text{I}^{+}]$ between the concentration of Cs and I vacancies will determine the negative/positive charge of the surface. The calculated formation energies of $V_\text{Cs}^{-}$ and $V_\text{I}^{+}$ vacancies at the surface are very close in the $p$-type region (Fig.~\ref{fig-def-ene}a). Thus, any subtle changes in the chemical potential of these two species can shift the balance. The experimental study of  \citet{Kim_JPCC_123_2019} supports the dominance of iodine vacancies at the (100) surface of \ch{MAPbI3}. Figure~S11 in the \gls{SI} section shows effect of the periodic slab electrostatic correction on the formation energy of charged defects using the tetragonal phase as an example. This correction does not change the overall trend.

\begin{figure}
  \includegraphics{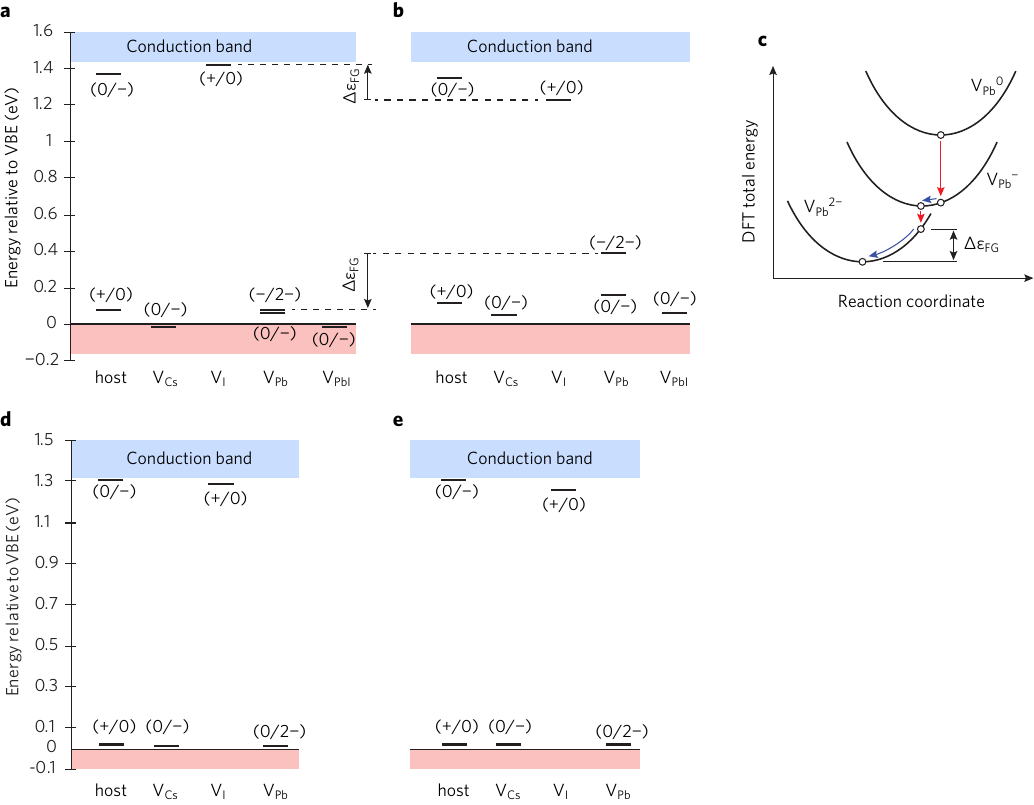}
  \caption{Transition energy levels for vacancies in cubic \ch{CsPbI3} at the surface (a,b) and in the bulk (d,e). Transition levels of the host without defects indicate the position of the perceived band edges. The energy levels are shown in a thermodynamic limit (a,d), meaning unconstrained reaction coordinates, and also constrained to vertical transitions (b,e). Values in brackets indicate the charge state. The configuration energy diagram (c) schematically shows the evolution of the \gls{DFT} total energy of the Pb surface vacancy at different charge states as a function of a reaction coordinate (atomic positions). Red arrows represent vertical electronic transitions, while blue arrows correspond to non-radiative energy dissipation processes. $\Delta\varepsilon_{\text{FG}}$ indicates a Franck-Condon shift, which is most prominent for the transition energy levels $\varepsilon(V_\text{Pb}^{-/2-})$ and $\varepsilon(V_\text{I}^{+/0})$ at the surface. The results are shown without \gls{SOC} to compensate for the \gls{DFT} band gap error. }
  \label{fig-transition-energy-levels}
\end{figure}

So far, we have discussed thermodynamic transition energy levels. When non-radiative transitions occur through deep energy levels, atomic positions do not have time to relax during the fast electronic transition (e.g., the capture of an electron from the \gls{CBE} by a deep trap). As a result, the thermodynamic and optical transition levels will differ by the value of a Franck-Condon shift, which can be as high as 0.5~eV for a carbon deep acceptor \ch{C_N^{0/-}} in GaN \cite{Lyons_APL_97_2010}. Signature of a large Franck-Condon shift would be a significant change in the \gls{DFT} total energy of the defect model, which occurs during structural relaxation in a charged state when starting from a \textit{fully relaxed} geometry of the neutral defect, indicating a large Franck-Condon shift. To assess the total energy of charged states without additional structural relaxation, we performed calculations of transition energy levels of defects using the fully relaxed geometry in the neutral state, and compared the results to when additional structural relaxation was included (Fig.~\ref{fig-transition-energy-levels}). In bulk, there are no major differences between transition energy levels calculated with and without relaxation of charged states. However, two energy levels ($V_\text{Pb}^{-/2-}$ and $V_\text{I}^{+/0}$) stand out from the others when calculated at the surface. The thermodynamic energy level of the shallow acceptor $V_\text{Pb}^{-/2-}$ becomes a \textit{deep} acceptor for optical electronic transitions, likely due to the loosely bound iodine at the surface (see \gls{SI} Fig.~S5b). The configuration energy diagram (Fig.~\ref{fig-transition-energy-levels}c) illustrates transitions between successive charge states, involving electronic transitions and non-radiative energy dissipation. It can be speculated that defects with a large Franck-Condon shift are likely to become centers of non-radiative recombination. However, it should be noted that the formation energy of the alternative $V_\text{PbI}$ divacancy at the surface is lower than that of $V_\text{Pb}$ in the $p$-type region (Fig.~\ref{fig-def-ene}). The energy level $\varepsilon(V_\text{PbI}^{0/-})$ is shallow and the Franck-Condon shift is rather small (Fig.~\ref{fig-transition-energy-levels}) making $V_\text{PbI}$ divacancy not a deep trap.

\subsection{Localization of defect states}

The wave function \gls{PR} is employed to quantify changes in its localization at band edges. The \gls{PR} of an eigenstate with the energy $E_{n,\mathbf{k}}$ is defined as \cite{Murphy_PRB_83_2011,Pashartis_PRA_7_2017}
\begin{equation}\label{eq:PR}
    \text{PR}(E_{n,\mathbf{k}}) = 
    \frac
    {
        \left(
            \sum_{\alpha=1}^{N} |\psi_{n,\mathbf{k},\alpha}|^2
        \right)^2
    }
    {
        \sum_{\alpha=1}^{N} |\psi_{n,\mathbf{k},\alpha}|^4
    },
\end{equation}
where $|\psi_{n,\mathbf{k},\alpha}|^2$ is the probability of finding an electron of the specific eigenstate within a sphere centered at the atomic site $\alpha$. The summation index $\alpha$ runs over all atomic sites $N$. The surface slab model has 288 atoms (63-Cs, 54-Pb, and 171-I), and the bulk model has 320 atoms (64-Cs, 64-Pb, and 192-I). The \gls{PR} has two extremes: (i) $\text{PR}=N$ when the wave function is fully delocalized (i.e., uniformly distributed among all $N$ atoms), and (ii) $\text{PR}=1$ if the wave function is localized on only one of the atoms. Since Cs atoms do not contribute to electronic states at the band edges of lead-halide perovskites, the upper limit of \gls{PR} will effectively become 226 and 256 for the slab and bulk models, respectively.

In defect-free structures, the \gls{VBE} has a higher \gls{PR} compared to \gls{CBE} (Table~\ref{tbl:PR}). The \gls{PR} of bulk \gls{VBE} is nearly identical to the number of Pb and I atoms in the model. This result can be explained by a nearly equal participation of both lead and iodine atoms in the \gls{VBE} state (Fig.~\ref{fig-structure}b), while the \gls{CBE} is dominated by lead with only a weak presence of iodine (Fig.~\ref{fig-structure}c). The \gls{PR} is reduced by 35\% in the slab \gls{VBE}, which is subject to a combined effect of quantum confinement and electric field due to surface reconstruction. The \gls{PR} of \gls{CBE} states in the bulk and slab with \gls{SOC} are nearly identical, taking into account the different number of atoms in the two models. This is ascribed to the quantum confinement and the electric field opposing each other. The delocalization due to \gls{SOC} (higher \gls{PR}) is noted in the defect-free slab model but not in bulk, suggesting that \gls{SOC} is an essential ingredient for a proper description of the spatial distribution of surface states.

\begin{table}
  \caption{Localization of electronic states at the band edges of cubic \ch{CsPbI3} with defects assessed by \gls{PR}. Values significantly lower than the reference without defects indicate localization. Values in the numerator/denominator refer to \gls{VBE}/\gls{CBE} states. The \gls{PR} of the \gls{VBE} state is underlined for $p$-type defects, and the \gls{CBE} state is underlined for $n$-type. $N$ is the total number of atoms in each model without defects.}
  \label{tbl:PR}
  \begin{tabular}{lccccc}
    \hline
    Defect  &  \multicolumn{2}{c}{Surface slab ($N=288$)}  && \multicolumn{2}{c}{Bulk ($N=320$)} \\
    \cline{2-3}\cline{5-6}
    & without \gls{SOC} & with \gls{SOC} &&  without \gls{SOC} & with \gls{SOC} \\
    \hline
    no defect (ref.) & 129/65 & 141/85 && 240/98 & 246/90 \\
    \hline
    $V_{\text{Cs}}$ & \underline{131}/31 & \underline{142}/71 && \underline{236}/96 & \underline{242}/91 \\
    $V_{\text{Pb}}$ & \underline{122}/30 & \underline{136}/67 && \underline{234}/88 & \underline{240}/88 \\
    $V_{\text{I}}$ & 125/\underline{46} & 116/\underline{67} && 233/\underline{55} & 234/\underline{91} \\
    $V_{\text{PbI}}$ & \underline{122}/29 & \underline{135}/69 && --- & --- \\
    \hline
  \end{tabular}
\end{table}

The $p$-type vacancies ($V_{\text{Cs}}$, $V_{\text{Pb}}$, and $V_{\text{PbI}}$) leave the \gls{PR} of the \gls{VBE} almost unchanged in the slab and in the bulk models (see Table~\ref{tbl:PR} as well as \gls{SI} Figs.~S6a,b,d and S7a,b,d for the spatial distribution of $|\psi_{\text{VBE}}(\mathbf{r})|^2$ in cubic and tetragonal models, respectively). The $n$-type defect ($V_{\text{I}}$) exhibits stronger localization of the \gls{CBE} state when \gls{SOC} is omitted, while the \gls{PR} of the \gls{CBE} is recovered back to the defect-free level with \gls{SOC} in the cubic structure (see \gls{SI} Fig.~S6c), indicating its importance for a proper description of the spatial distribution of states associated with $n$-type defects. In the tetragonal phase with the iodine vacancy at the surface, the \gls{CBE} states remain confined at the surface despite \gls{SOC} (see \gls{SI} Fig.~S7c); it does not create a deep trap, though, as evident from the shallowness of the transition energy level $\varepsilon(V_{\text{I}}^{+/0})$ (Table~\ref{tbl:Transition-ene-levels}).

\section{Conclusions}

Motivated by the importance of interfaces in layered perovskites, nanoparticles, and polycrystalline thin films, we characterized the formation of vacancies at the surface of a slab model and contrasted the results to the bulk of lead-halide perovskites using cubic and tetragonal \ch{CsPbI3} as model structures. Our slab model is equivalent to the 2D Ruddlesden-Popper structure with six layers. Thus, we anticipate that the results will also be transferable to layered perovskites.

The free \ch{CsI}-terminated surface does not act as a trap for free carriers, since band edge states are not localized at the surface. The surface reconstruction creates an electric field that attracts electrons towards the surface but not strong enough to localize electrons at the surface. There is no significant perturbation of \gls{VBE} states due to $p$-type vacancies ($V_{\text{Cs}}$, $V_{\text{Pb}}$, and $V_{\text{PbI}}$) both at the surface and in the bulk. However, there is a notable attractive potential for electrons towards $V_{\text{I}}$ at the surface, especially in the case of the tetragonal phase. Formation energies of vacancies were evaluated with chemical potentials of species tailored to an aqueous solution. The chemical potentials were confined to a rather small region, which eliminated the largest source of uncertainty in previous calculations of defects' formation energies. The surface favors the formation of Cs vacancies, while Pb vacancies are more favorable in the bulk. The formation energy of iodine vacancies at the surface is very similar to that in the bulk. Lead-iodine divacancies ($V_\text{PbI}^{-}$) are expected to dominate over lead-only vacancies at the surfaces. The analysis of charged defects at the surface suggests pinning of the Fermi energy at the \gls{VBE} and segregation of $V_\text{Cs}^{-}$ and $V_\text{I}^{+}$ vacancies (both have a low formation energy of 0.25$-$0.4~eV). The predominant vacancies in the bulk are $V_\text{Pb}^{2-}$ and $V_\text{I}^{+}$ with a formation energy of ca. 0.5~eV. The chemical tolerance of our (defect) formation energy calculations is about 0.2~eV at most. Energy levels of dominant vacancies at the surface remain close to the band edges and do not form detrimental deep traps within the band gap. These states fall into the category of delocalized host states perturbed by a screened Coulomb potential and follow band edges when the band gap is renormalized by \gls{SOC}. There are no major differences between energy levels calculated with and without additional structural relaxation caused by changes in occupancy of electronic levels when transitioning from one charge state to another for dominant defects. The negligible structural relaxation between $q \pm 1$ states indicates a rather small Franck-Condon shift. Defects with a large Frank-Condon shift are likely to become centers of non-radiative recombination. These results explain the high optoelectronic performance of two-dimensional structures, nanoparticles, and polycrystalline thin films of lead-halide perovskites in spite of the abundance of interfaces in these materials.

Finally, it should be noted that there is no unique suggestion regarding the omission or inclusion of \gls{SOC} when combined with \gls{DFT}-\gls{PBE} for analysis of defects in this material system. The relativistic electronic structure of \gls{CBE} in the slab is more resilient to the formation of surface states due to the mixing of Pb-$p_{x,y,z}$ orbitals by the \gls{SOC} Hamiltonian. Therefore, it is recommended to include \gls{SOC} in the analysis of the spatial distribution of the wave function for \gls{CBE} states, also for $n$-type defects. The formation energy of iodine vacancies is greatly underestimated (by ca. 1~eV) in a calculation at the \gls{DFT}-\gls{PBE} level of theory with \gls{SOC} due to its large band gap error. It is, therefore, imperative not to combine \gls{PBE} with \gls{SOC} when computing the formation energy of $n$-type defects in this material system.

\begin{acknowledgement}

Authors are thankful to Camille Latouche and Th{\'e}o Cavignac (University of Nantes) for a discussion about PyDEF~2. Calculations were performed using the Digital Research Alliance of Canada (Compute Canada) infrastructure supported by the Canada Foundation for Innovation under the John R. Evans Leaders Fund. X.R. acknowledges GENCI for granting access to the HPC resources of [TGCC/CINES/IDRIS] under the allocation 2022-A0130907682. 

\end{acknowledgement}

\subsection*{Supplementary information}

Supplementary information (a PDF file) is available free of charge from the publisher. It contains illustrations of the spatial distribution of the \gls{VBE} and \gls{CBE} wave function probability densities $|\psi(\mathbf{r})|^2$ for the slab model calculated with and without defects at different levels of theory. The local structure of vacancies at the surface is shown. We have also included details on calculation of the electrostatic charge correction for periodic slabs. The equivalent of Fig.~\ref{fig-def-ene} for the tetragonal phase is presented with and without the electrostatic charge correction for periodic slabs. The following references \cite{Freysoldt_PRB_97_2018, Freysoldt_PRL_102_2009, Srikanth_PCCP_22_2020, Makov_PRB_51_1995} were given credit in \gls{SI}.

\subsection*{Data availability}

\Gls{VASP} structure and output files are available from the Zenodo repository \cite{Zenodo_10.5281/zenodo.8329348}. The repository also includes the output of PyDEF~2 with an analysis of defect characteristics along with the PyDEF~2 main input file used in data processing. The modified version (fork) of PyDEF~2 code needed to reproduce this work is available from GitHub \cite{git_PyDEF2_fork}. \Gls{VASP} structure files can be visualized using the Vesta program \cite{Momma_JAC_44_2011}.

\bibliography{bibliography}

\clearpage

\renewcommand\thesection{S\arabic{section}}
\renewcommand\thefigure{S\arabic{figure}}
\renewcommand\thetable{S\arabic{table}}
\renewcommand\theequation{S\arabic{equation}}
\renewcommand\thepage{S\arabic{page}}
\setcounter{section}{0}
\setcounter{figure}{0}
\setcounter{table}{0}
\setcounter{equation}{0}
\setcounter{page}{1}

\section*{Supplementary information for publication}

Title: \textbf{Defect Tolerance of Lead-Halide Perovskite (100) Surface Relative to Bulk: Band Bending, Surface States, and Characteristics of Vacancies}

\noindent Authors: \textit{Oleg Rubel}$^\dag$ and \textit{Xavier Rocquefelte}$^\ddag$

\noindent Affiliations: $^\dag$Department of Materials Science and Engineering, McMaster University, 1280 Main Street West, Hamilton, Ontario L8S 4L8, Canada; $^\ddag$Univ Rennes, CNRS, ISCR (Institut des Sciences Chimiques de Rennes) UMR 6226, 263 Av. G{\'e}n{\'e}ral Leclerc, 35700 Rennes, France.

\vspace{18pt}

\begin{figure}
  \includegraphics{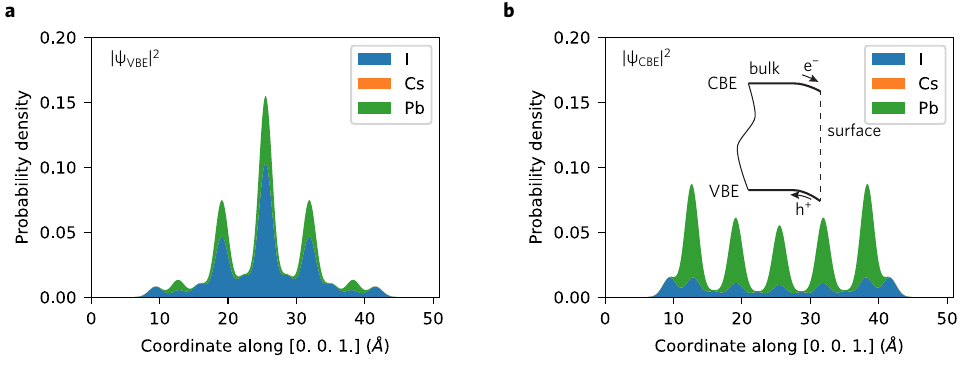}
  \caption{Wave function distribution in a slab of tetragonal \ch{CsPbI3} calculated with \gls{PBE}-\gls{SOC}: (a,b) band edges (\gls{VBE} and \gls{CBE}) with surface reconstruction calculated at \gls{HSE06} level of theory, (c,d) band edges without surface reconstruction calculated at \gls{PBE} level of theory.}
  \label{fig-SI-psink-surf-SOC-tet}
\end{figure}

\begin{figure}
  \includegraphics{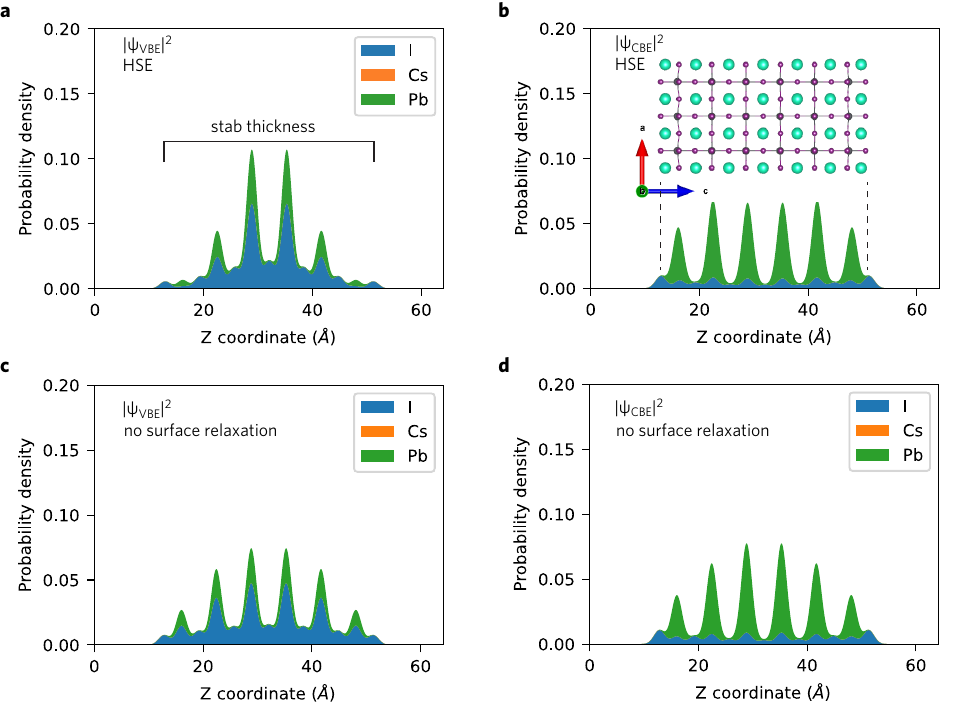}
  \caption{Wave function distribution in a slab of cubic \ch{CsPbI3} calculated with \gls{SOC}: (a,b) band edges (\gls{VBE} and \gls{CBE}) with surface reconstruction calculated at \gls{HSE06} level of theory, (c,d) band edges without surface reconstruction calculated at \gls{PBE} level of theory.}
  \label{fig-SI-psink-surf-SOC}
\end{figure}

\begin{figure}
  \includegraphics{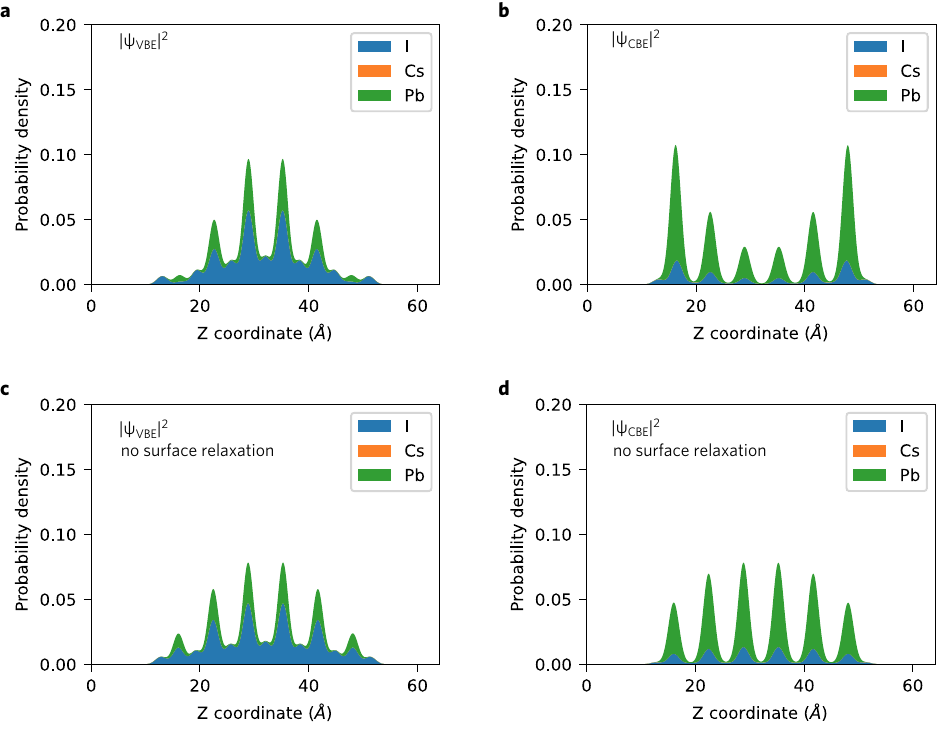}
  \caption{Wave function distribution in a slab of cubic \ch{CsPbI3} calculated \textbf{without} \gls{SOC} at \gls{PBE} level of theory: (a,b) band edges (\gls{VBE} and \gls{CBE}) with surface reconstruction, (c,d) band edges without surface reconstruction.}
  \label{fig-SI-psink-surf-noSOC}
\end{figure}

\begin{figure}
  \includegraphics[scale=0.9]{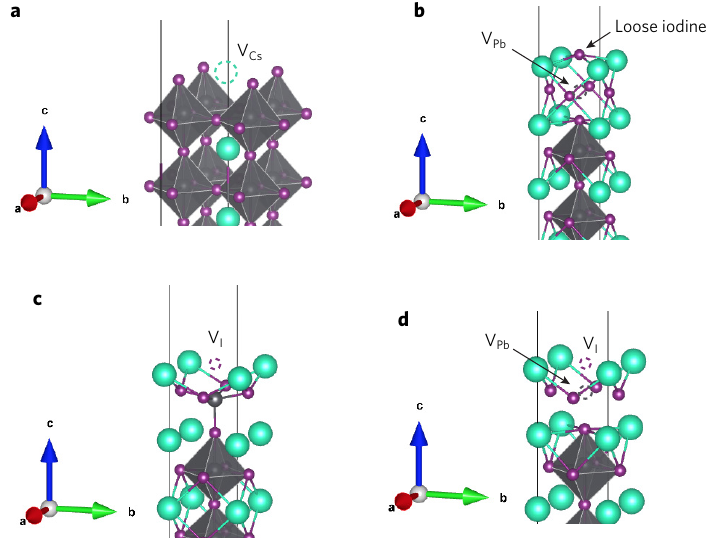}
  \caption{Local structure of surface defects: (a) cesium vacancy, (b) lead vacancy, (c) iodine vacancy, (d) lead-iodine divacancy. Only a small fragment in the immediate vicinity of defects is shown.}
  \label{fig-SI-struct-defects-surf}
\end{figure}

\begin{figure}
  \includegraphics{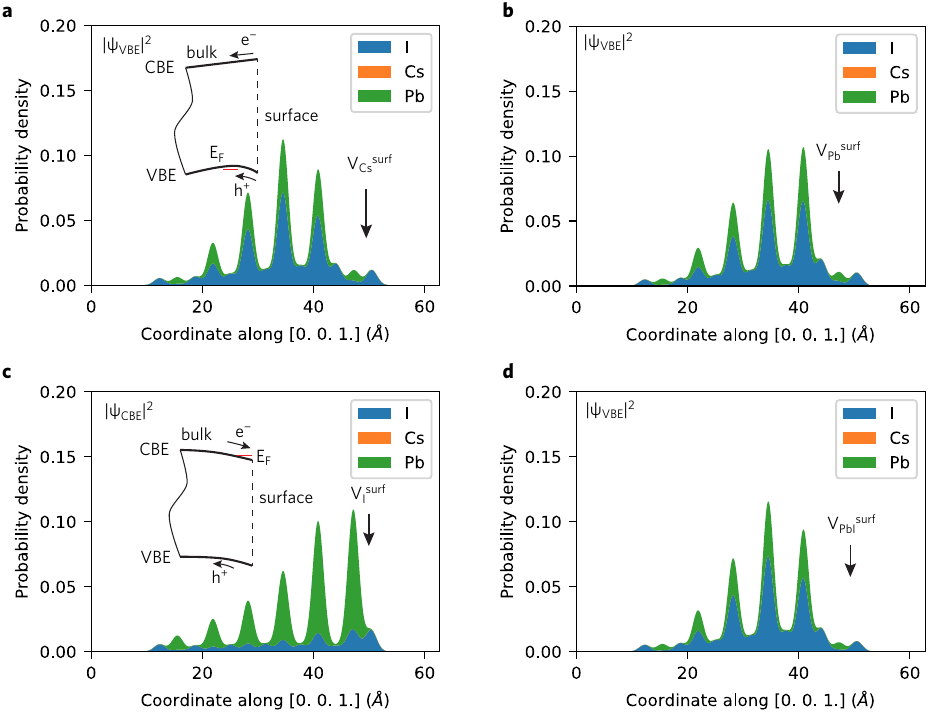}
  \caption{Wave function distribution at band edges (\gls{VBE} and \gls{CBE}) in a slab of cubic \ch{CsPbI3} with surface defects: (a) $V_{\text{Cs}}$, (b) $V_{\text{Pb}}$, (c) $V_{\text{I}}$, and (d) $V_{\text{PbI}}$. The location of defects is marked with arrows. The calculations include \gls{SOC}. Inset shows a schematic band diagram.}
  \label{fig-SI-psink-surf-defects-SOC}
\end{figure}

\begin{figure}
  \includegraphics{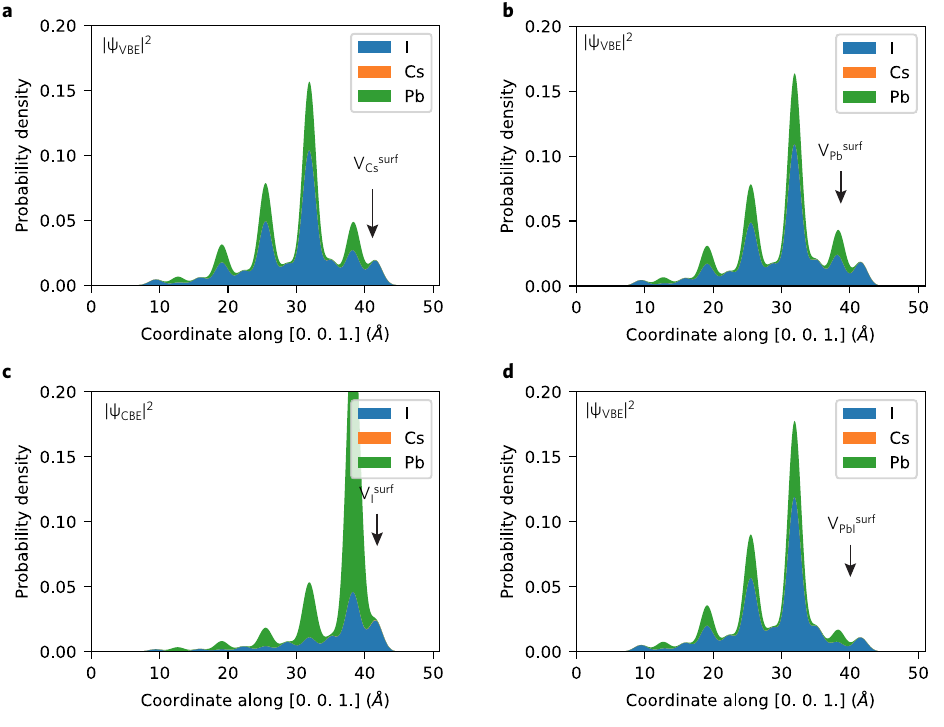}
  \caption{Wave function distribution at band edges (\gls{VBE} and \gls{CBE}) in a slab of tetragonal \ch{CsPbI3} with surface defects: (a) $V_{\text{Cs}}$, (b) $V_{\text{Pb}}$, (c) $V_{\text{I}}$, and (d) $V_{\text{PbI}}$. The location of defects is marked with arrows. The calculations include \gls{SOC}.}
  \label{fig-SI-psink-surf-defects-SOC-tet}
\end{figure}

Figures~\ref{fig-SI-psink-surf-SOC-tet}$-$\ref{fig-SI-psink-surf-noSOC}, \ref{fig-SI-psink-surf-defects-SOC}, and \ref{fig-SI-psink-surf-defects-SOC-tet} show the atom-resolved \gls{DFT} wave functions $|\psi_{n,\mathbf{k},\alpha}(z)|^2$ for the \gls{VBE} and \gls{CBE},  normalized such that $\int \sum_{\alpha} |\psi_{n,\mathbf{k},\alpha}(z)|^2~dz = 1$. Here, $n$ is the band index, $\mathbf{k}$ is the electron wave vector, and $\alpha$ is the atom label that runs over all atoms. The local structure of vacancies at the surface is  shown in Fig.~\ref{fig-SI-struct-defects-surf}. Additional tests were performed to confirm that the results in Fig.~\ref{fig-SI-psink-surf-defects-SOC} remain valid when a dipole correction is activated (\gls{VASP} tags \texttt{LDIPOL=.TRUE.}, \texttt{IDIPOL=3}, \texttt{DIPOL=0.5 0.5 0.5}).

\begin{figure}
  \includegraphics{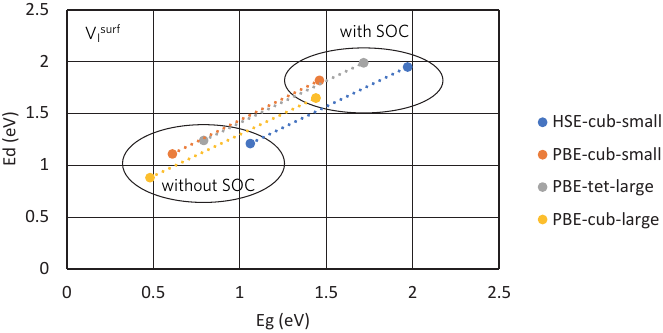}
  \caption{Formation energy of the neutral iodine vacancy at the slab surface correlated with the calculated band gap. The `small' , `large' cubic and `large' tetragonal slab models refer to the sizes $2 \times 2 \times 3$, $3 \times 3 \times 6$, and $2\sqrt{2} \times 2\sqrt{2} \times 5$, respectively.}
  \label{fig-SI-Ed-V_I-surf-vs-Eg}
\end{figure}

Figure~\ref{fig-SI-Ed-V_I-surf-vs-Eg} shows the correlation between the formation energy of a neutral iodine vacancy at a slab surface and the calculated band gap. Therefore, it is important to select an appropriate structural model in combination with the appropriate level of theory to obtain reliable results.

\begin{figure}
  \includegraphics{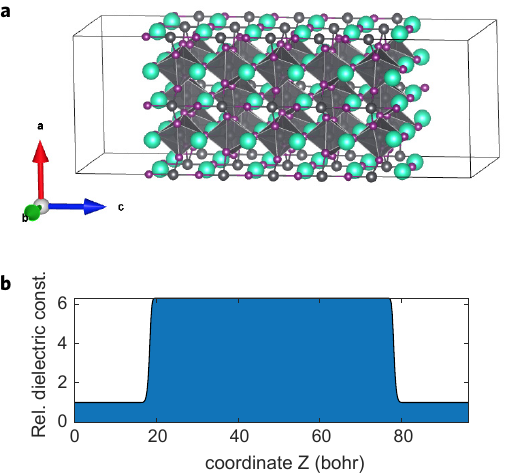}
  \caption{Effective dielectric model used in the evaluation of a correction for periodic charged slabs of tetragonal \ch{CsPbI3}: (a) Periodic slab structure. (b) Dielectric profile.}
  \label{fig-SI-dielectric-plus-model}
\end{figure}

The periodic electrostatic energy for charged slab models
$U_\text{per}$ was calculated using the \texttt{sxdefectalign2d} package \cite{Freysoldt_PRB_97_2018, Freysoldt_PRL_102_2009} version~1.1. The model excess charge density $\delta\rho(z)$ was approximated by a superposition of Gaussian functions. (Note that \texttt{sxdefectalign2d} uses a charge convention with a negative sign for positive charge.) The dielectric profile was approximated by a rectangular shape (Fig.~\ref{fig-SI-dielectric-plus-model}), assuming an isotropic relative dielectric constant of 6.3 \cite{Srikanth_PCCP_22_2020} for the slab materials. A cut-off energy of 40~Ry was selected to match the discretization of the Fourier space along the $z$-axis to that in \gls{VASP} \texttt{LOCPOT} files. A large in-plane Gaussian broadening of $\beta_{\parallel}=200$~bohr in the lateral direction was selected to represent charge delocalization. The model potential $\delta\phi(z)$ was calculated by \texttt{sxdefectalign2d} via solving a periodic Poisson's equation.

The listing below shows a sample of \texttt{sxdefectalign2d} input file \texttt{system.sx} for a charged iodine vacancy (without \gls{SOC}). The total charge $\sum Q= -1$ implies the positively charged defect (deficiency of $1e$). The excess charge is approximated by 11 Gaussian functions, each with a spread of 3~bohr along the $z$ axis:
\begin{verbatim}
slab {
    fromZ = 17.827539465159465;
    toZ = 78.59547174181203;
    epsilon = 6.3;
}
isolated {
    fromZ = 48.21150560348575;
    toZ = 87.47975407924827;
}
charge {
    posZ = 20.856600667379407;
    Q = -0.09090909090909091;
    betaZ = 3;
    betaPara = 200;
}
charge {
    posZ = 21.082790309866194;
    Q = -0.09090909090909091;
    betaZ = 3;
    betaPara = 200;
}
charge {
    posZ = 21.65211298378684;
    Q = -0.09090909090909091;
    betaZ = 3;
    betaPara = 200;
}
charge {
    posZ = 57.7869527991119;
    Q = -0.09090909090909091;
    betaZ = 3;
    betaPara = 200;
}
charge {
    posZ = 75.44635482632142;
    Q = -0.09090909090909091;
    betaZ = 3;
    betaPara = 200;
}
charge {
    posZ = 74.2780842525695;
    Q = -0.09090909090909091;
    betaZ = 3;
    betaPara = 200;
}
charge {
    posZ = 74.86934094464065;
    Q = -0.09090909090909091;
    betaZ = 3;
    betaPara = 200;
}
charge {
    posZ = 75.44864738923125;
    Q = -0.09090909090909091;
    betaZ = 3;
    betaPara = 200;
}
charge {
    posZ = 75.41614185661757;
    Q = -0.09090909090909091;
    betaZ = 3;
    betaPara = 200;
}
charge {
    posZ = 75.2970505657695;
    Q = -0.09090909090909091;
    betaZ = 3;
    betaPara = 200;
}
charge {
    posZ = 75.50571889448972;
    Q = -0.09090909090909091;
    betaZ = 3;
    betaPara = 200;
}
\end{verbatim}
The code is executed using the command\\
\verb|sxdefectalign2d --vref LOCPOT --vdef LOCPOTchrg_m1 --vasp --ecut 40|\\
The sample output is listed below:
\begin{verbatim}
VASP mesh: 140 x 140 x 384
cell defect = [a1={33.0657,0,0},a2={0,33.0657,0},a3={0,0,96.364}]
VASP mesh: 140 x 140 x 384
cell bulk = [a1={33.0657,0,0},a2={0,33.0657,0},a3={0,0,96.364}]
Periodic supercell taken from LOCPOTchrg_p1
[a1={33.0657,0,0},a2={0,33.0657,0},a3={0,0,96.364}]
Cutoff = 40 Ry
N = 396
--- Periodic
Q=-1
short-range potential with averaging width = 2 bohr
Slope bottom (@z=48.182)= 0.00456743 eV/bohr, value = -0.00937142 eV
Slope top (@z=87.6037)   = 0.00198296 eV/bohr, value = -0.070616 eV
---
--- Isolated
Isolated from 48.1820175594 to 87.6036682897 (162 points)
Q=-0.727204072125
Interface: 78.5954717418
zL = 48.1820175594 eps = 6.3
E-field dependence left:  zeff = 73.9587635068
zR = 87.6036682897 eps = 1
E-field dependence right: zeff = 77.8594863077
---
isolated energy = 0.00853691299722 eV
periodic energy = 0.108846350821 eV
iso - periodic energy = -0.100309437824 eV
\end{verbatim}
We are interested in the periodic energy $U_\text{per}=0.11$~eV. This energy needs to be \textit{subtracted} from the formation energy of a charged defect. By including this correction, we obtain the transition energy level $\varepsilon(V_{\text{I}}^{+/0})=E_{\text{CBE}}-0.08$~eV instead of $E_{\text{CBE}}-0.19$~eV reported in Table~\ref{tbl:Transition-ene-levels} for the tetragonal slab model without \gls{SOC}. The isolated energy is not exactly zero due to the finite size of $\beta_{\parallel}$, but it becomes zero in the limit of $\beta_{\parallel} \rightarrow \infty$. The output file \texttt{vline-eV.dat} contains the model potential $\delta\phi$, the \gls{DFT} potential $\Delta V_\text{DFT}$, and their difference. We used an auxiliary Python code to optimize positions of the Gaussian functions and minimize the residual between $\delta\phi$ and $\Delta V_\text{DFT}$. Results for other charged defects in the $2\sqrt{2}\times 2\sqrt{2}\times 5$ tetragonal slab are presented in Figs.~\ref{fig-SI-noSOC-sxdefectalign2d} and \ref{fig-SI-SOC-sxdefectalign2d} (without and with \gls{SOC}, respectively).

\begin{figure}
    \includegraphics{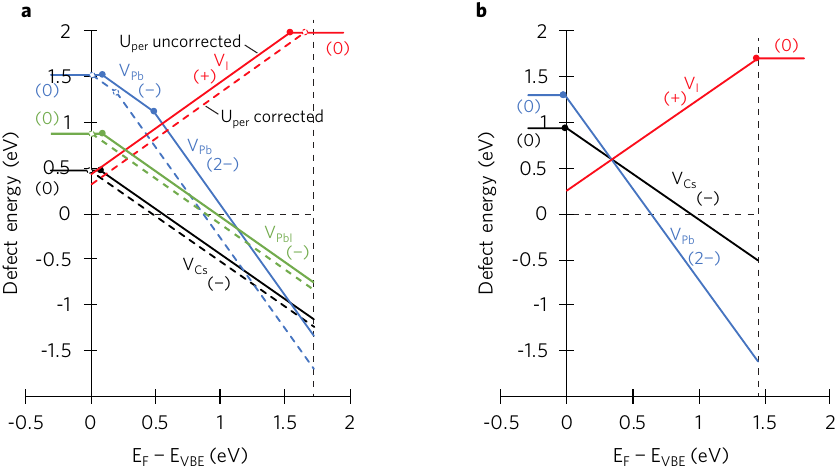}
    \caption{Formation energy of vacancies at the surface (a) and in the bulk (b) of tetragonal \ch{CsPbI3} as a function of the Fermi energy ($E_{\text{F}}$). Values in brackets represent the charge state. The vertical dashed lines mark the location of the \gls{CBE}. The chemical potential of elements is tuned to reflect solution-processed synthesis conditions. Results are shown without \gls{SOC} to compensate for the \gls{DFT} band gap error. Panel (a) presents data with and without the electrostatic correction $U_\text{per}$ for charged defects in periodic slabs.}
    \label{fig-SI-def-ene-tet}
\end{figure}

If the periodic slab correction (Fig.~\ref{fig-SI-noSOC-sxdefectalign2d}) was applied to energies of charged defects, all energies of defects with the charge $\pm 1$ would be shifted down by ca.~0.05$-$0.1~eV as shown in Fig.~\ref{fig-SI-def-ene-tet} for the tetragonal phase. The correction magnitude increases to 0.33$-$0.36~eV for the formation energy of $V_\text{Pb}^{2-}$ (Figs.~\ref{fig-SI-noSOC-sxdefectalign2d} and \ref{fig-SI-SOC-sxdefectalign2d}), which follows the $q^2$ scaling \cite{Makov_PRB_51_1995}.

\begin{figure}
    no defect $(q=+1)$, $U_\text{per}=0.06$~eV\quad\quad\quad\quad no defect $(q=-1)$, $U_\text{per}=0.07$~eV\\
    \includegraphics{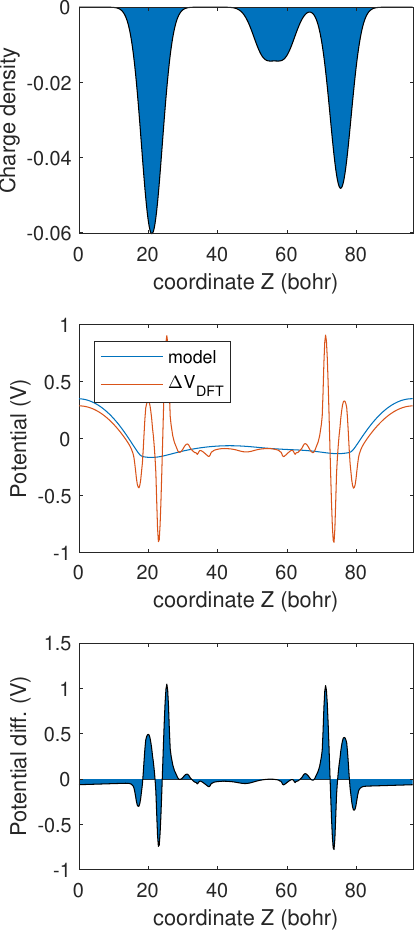}\quad\quad\quad\includegraphics{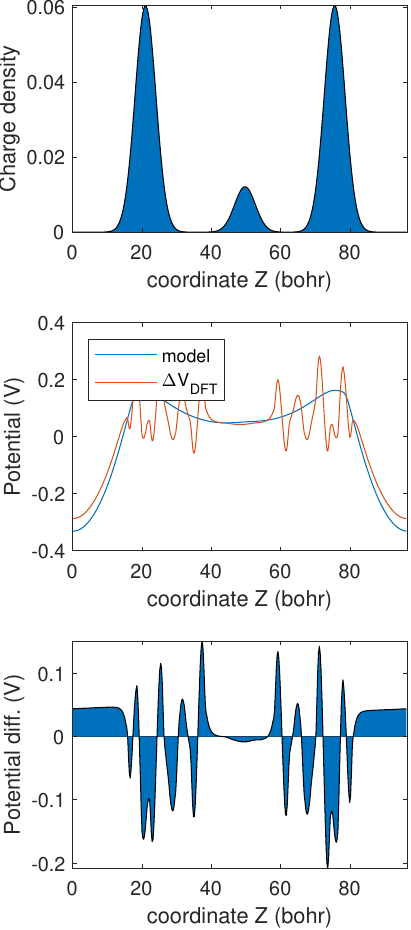}
    \caption{Output of \texttt{sxdefectalign2d} for an electrostatic model used in the evaluation of a correction for periodic charged slabs of tetragonal \ch{CsPbI3} with and without defects and the excess charge $q$ delocalized in the lateral plane: (top) Linear charge density distribution $\delta\rho(z)$. (middle) Model potential profile $\delta\phi(z)$ and the DFT potential $\Delta V_\text{DFT}(z)$. (bottom) Potential difference $\Delta V_\text{DFT} - \delta\phi$. \gls{SOC} is not included.}
    \label{fig-SI-noSOC-sxdefectalign2d}
\end{figure}

\begin{figure}
    \ContinuedFloat
    Cs vacancy $(q=-1)$, $U_\text{per}=0.08$~eV\quad\quad\quad\quad I vacancy $(q=+1)$, $U_\text{per}=0.11$~eV\\
    \includegraphics{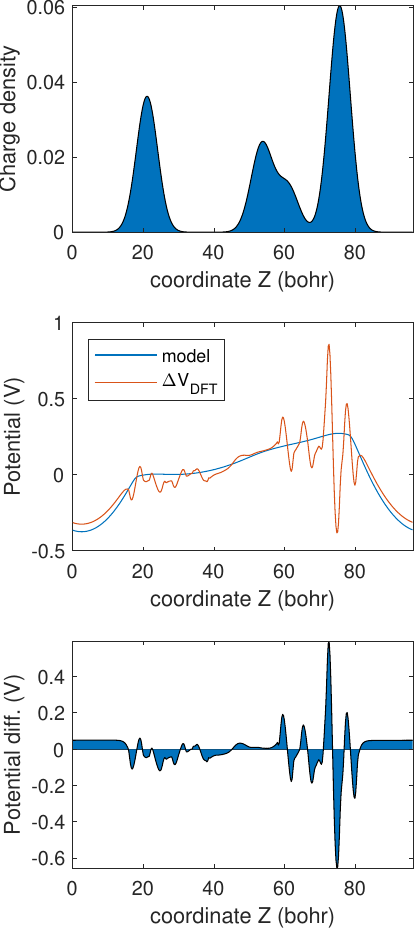}\quad\quad\quad\includegraphics{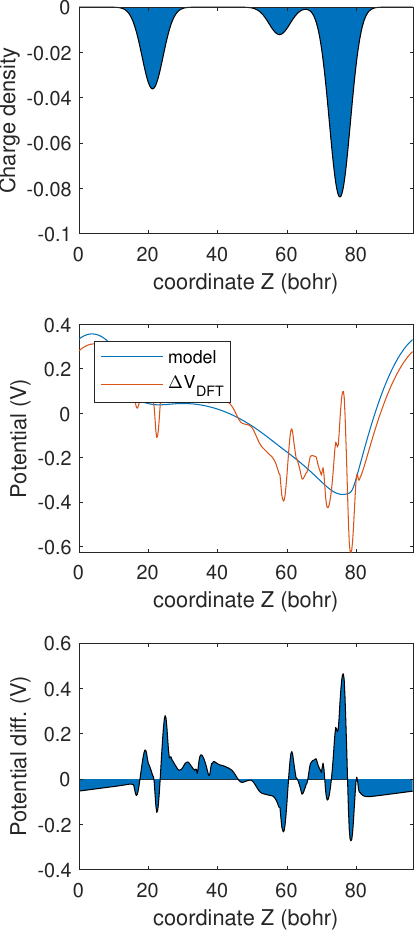}
    \caption{(continued)}
\end{figure}

\begin{figure}
    \ContinuedFloat
    Pb vacancy $(q=-1)$, $U_\text{per}=0.08$~eV\quad\quad\quad\quad Pb vacancy $(q=-2)$, $U_\text{per}=0.36$~eV\\
    \includegraphics{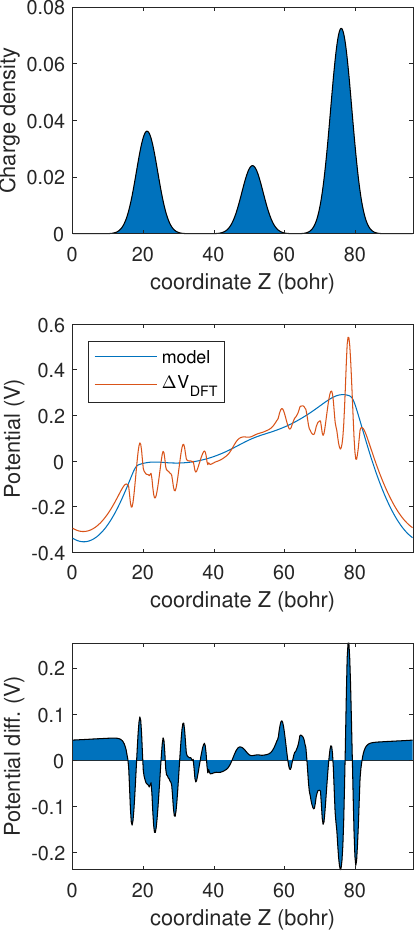}\quad\quad\quad\includegraphics{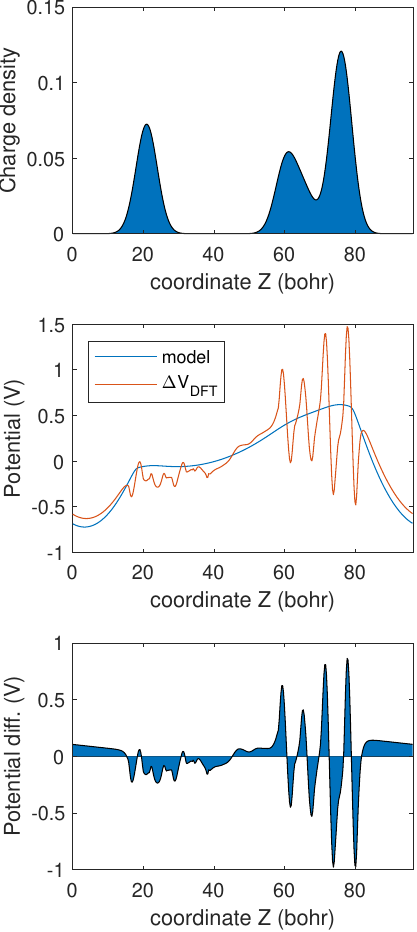}
    \caption{(continued)}
\end{figure}

\begin{figure}
    \ContinuedFloat
    PbI divacancy $(q=-1)$, $U_\text{per}=0.08$~eV\\
    \includegraphics{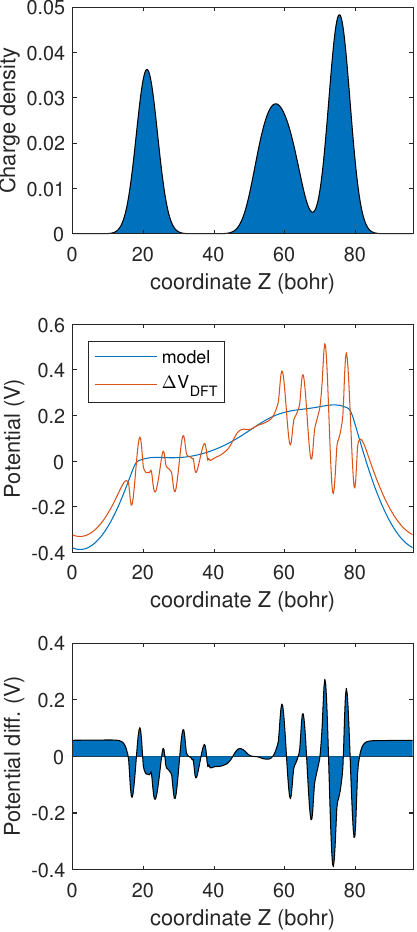}
    \caption{(continued)}
\end{figure}

\begin{figure}
    no defect $(q=+1)$, $U_\text{per}=0.06$~eV\quad\quad\quad\quad no defect $(q=-1)$, $U_\text{per}=0.06$~eV\\
    \includegraphics{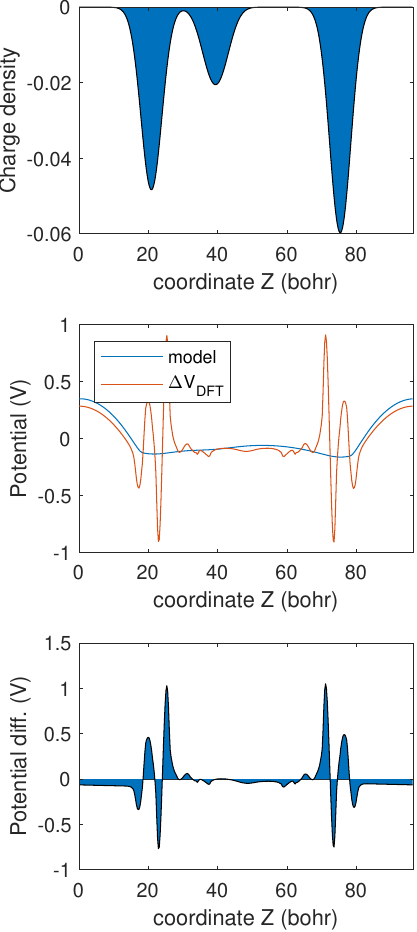}\quad\quad\quad\includegraphics{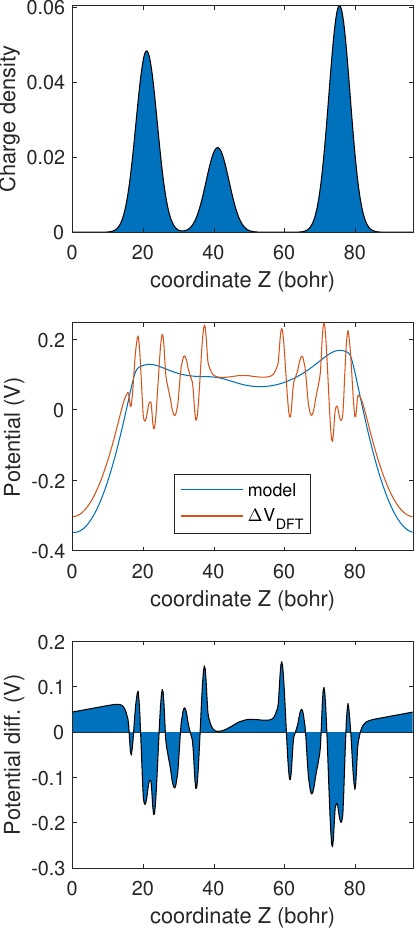}
    \caption{Output of \texttt{sxdefectalign2d} for an electrostatic model used in the evaluation of a correction for periodic charged slabs of tetragonal \ch{CsPbI3} with and without defects and the excess charge $q$ delocalized in the lateral plane: (top) Linear charge density distribution $\delta\rho(z)$. (middle) Model potential profile $\delta\phi(z)$ and the DFT potential $\Delta V_\text{DFT}(z)$. (bottom) Potential difference $\Delta V_\text{DFT} - \delta\phi$. \gls{SOC} is included.}
    \label{fig-SI-SOC-sxdefectalign2d}
\end{figure}

\begin{figure}
    \ContinuedFloat
    Cs vacancy $(q=-1)$, $U_\text{per}=0.07$~eV\quad\quad\quad\quad I vacancy $(q=+1)$, $U_\text{per}=0.06$~eV\\
    \includegraphics{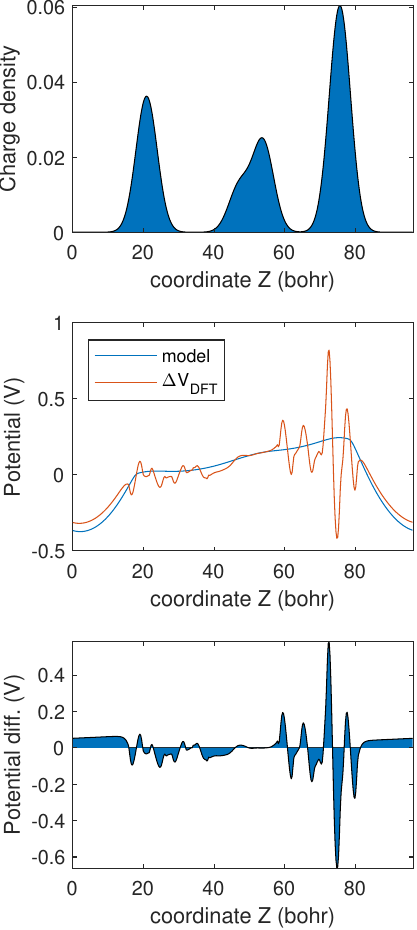}\quad\quad\quad\includegraphics{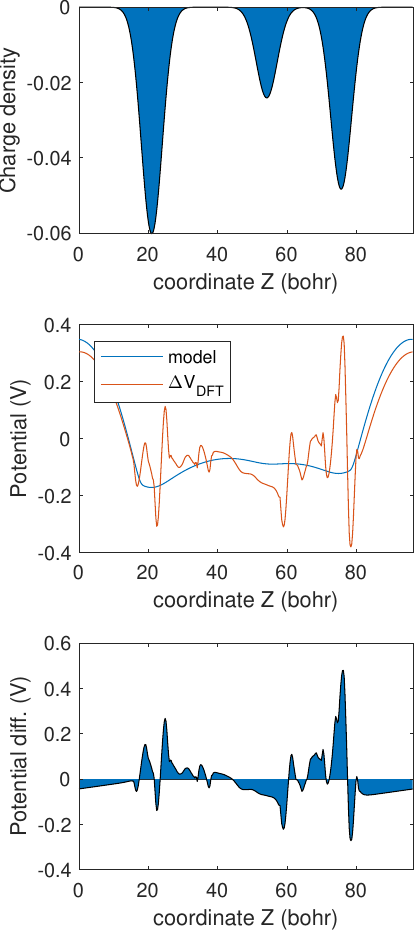}
    \caption{(continued)}
\end{figure}

\begin{figure}
    \ContinuedFloat
    Pb vacancy $(q=-1)$, $U_\text{per}=0.07$~eV\quad\quad\quad\quad Pb vacancy $(q=-2)$, $U_\text{per}=0.33$~eV\\
    \includegraphics{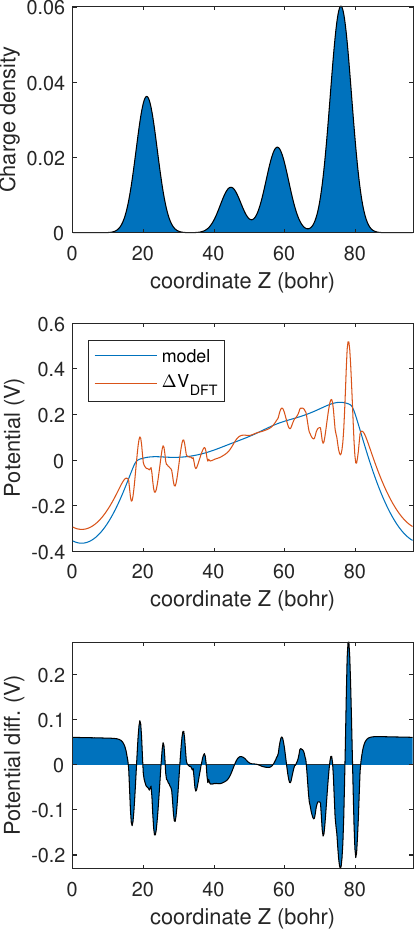}\quad\quad\quad\includegraphics{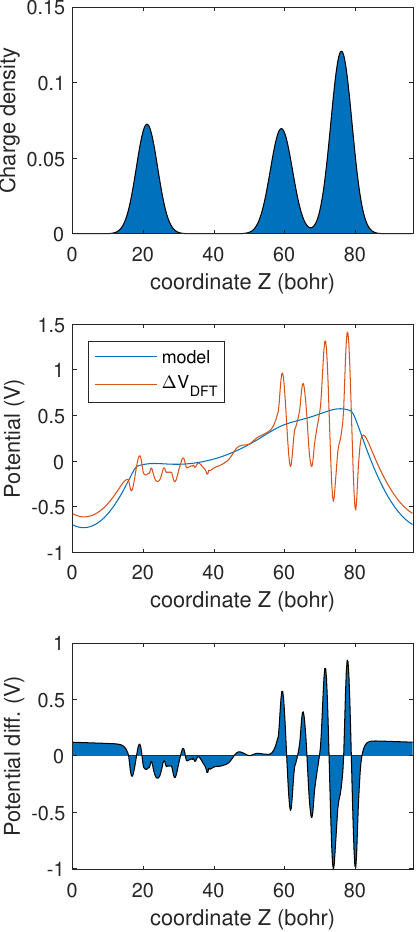}
    \caption{(continued)}
\end{figure}

\begin{figure}
    \ContinuedFloat
    PbI divacancy $(q=-1)$, $U_\text{per}=0.07$~eV\\
    \includegraphics{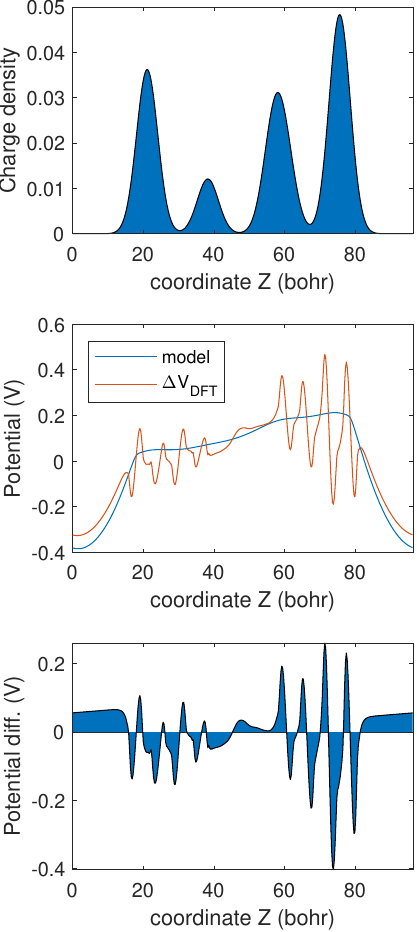}
    \caption{(continued)}
\end{figure}

\end{document}